\documentclass[journal]{vgtc}              

\onlineid{1175}

\vgtccategory{Research}

\vgtcpapertype{Representation \& Interaction}

\title{``Mapping What I Feel'': Understanding Affective Geovisualization Design Through the Lens of People-Place Relationships}

\author{Xingyu Lan*, Yutong Yang* and Yifan Wang
}

\authorfooter{
    \item 
        Xingyu Lan is with Fudan University.
        E-mail: xingyulan96@gmail.com.
    \item 
        Yutong Yang is with Shanghai Jiao Tong University.
  	E-mail: flora20@sjtu.edu.cn.
    
    \item
  	Yifan Wang is with Fudan University.
  	E-mail: wangyifanlea@gmail.com.
    \item
    *~Equal contributions.

}

\abstract{
Affective visualization design is an emerging research direction focused on communicating and influencing emotion through visualization. However, as revealed by previous research, this area is highly interdisciplinary and involves theories and practices from diverse fields and disciplines, thus awaiting analysis from more fine-grained angles. To address this need, this work focuses on a pioneering and relatively mature sub-area, affective geovisualization design, to further the research in this direction and provide more domain-specific insights. Through an analysis of a curated corpus of affective geovisualization designs using the Person-Process-Place (PPP) model from geographic theory, we derived a design taxonomy that characterizes a variety of methods for eliciting and enhancing emotions through geographic visualization. We also identified four underlying high-level design paradigms of affective geovisualization design (e.g., computational, anthropomorphic) that guide distinct approaches to linking geographic information with human experience. By extending existing affective visualization design frameworks with geographic specificity, we provide additional design examples, domain-specific analyses, and insights to guide future research and practices in this underexplored yet highly innovative domain.
}


\graphicspath{{figs/}{figures/}{pictures/}{images/}{./}} 

\usepackage{tabu}                      
\usepackage{booktabs}                  
\usepackage{lipsum}                    
\usepackage{mwe}                       

\usepackage{mathptmx}                  

\newcommand{\etal}{et~al.~} 
\newcommand{\ie}{i.e.,~}
\newcommand{\eg}{e.g.,~}
\newcommand{\ncorpus}{62 }

\usepackage{graphicx}
\usepackage{wrapfig}
\usepackage[T1]{fontenc}
\usepackage{amsmath}  
\usepackage{multirow}
\usepackage{tabularx}
\newcommand{\x}[1]{{\leavevmode\color{brown}{#1}}}
\usepackage[svgnames,x11names,table]{xcolor}
\usepackage{soul}
\usepackage{adjustbox}
\usepackage{rotating}
\usepackage[normalem]{ulem}

\definecolor{pblue}{HTML}{4E98C0}
\definecolor{ppink}{HTML}{B78296}
\definecolor{pyellow}{HTML}{AF967E}
\definecolor{pgreen}{HTML}{58ae98}
\newcommand{\person}[1]{\textcolor{ppink}{#1}}
\newcommand{\place}[1]{\textcolor{pblue}{#1}}
\newcommand{\processi}[1]{\textcolor{pyellow}{#1}}
\newcommand{\processii}[1]{\textcolor{pgreen}{#1}}

\begin{document}

\firstsection{Introduction}
\maketitle

In recent years, \textit{affective visualization design}, which uses visualization to communicate or influence emotion~\cite{lan2023affective}, has risen as a trendy research direction. This shift is driven by a growing recognition that visualization is not merely an objective representation of data but also integrates subjective factors at various stages of data processing and dissemination~\cite{lee2022affective}. Meanwhile, a series of user studies (\eg ~\cite{lan2021kineticharts,thudt2018self}) have found that affective visualization design makes visualizations more attractive, engaging, and memorable. Such designs have also been increasingly applied in public affairs to help spark attention and raise people's awareness of issues such as healthcare and social inequality~\cite{lan2022negative,boy2017showing}.
According to the survey by Lan~\etal~\cite{lan2023affective}, among various domains that have investigated this topic, the field of geography stands out as a pioneer.
Since the early 2000s, geographers have successively introduced concepts such as \textit{emotional geography}~\cite{anderson2001emotional} and \textit{affective geovisualization}~\cite{aitken2009into}, arguing that ``to neglect the emotions is to exclude a key set of relations through which lives are lived and societies made''~\cite{anderson2001emotional}.
In this context, many studies have been conducted to examine emotional factors in areas such as urban planning, human migration, and place attachment~\cite{bondi2016emotional,smith2009emotion}. 
The identified papers about affective visualization design in the field of geography even reached a small peak before computer science began to dominate this research direction~\cite{lan2023affective}. Given the pioneering and substantial accumulation in the field of geography, we argue that to further deepen the research on affective visualization design, geography is a sub-field that particularly deserves attention and specialized investigation.


On the other hand, geographic data has always been a vital research target in the visualization community. However, previous work has mostly focus on the visual analytics of geographic data (\eg analyzing users' movement patterns~\cite{chen2017vaud}, mining places and associated events from documents~\cite{cho2015vairoma}) or the effective design of geographic visualizations (\eg how to optimize the coloring of maps~\cite{harrower2003colorbrewer,golbiowska2020rainbow}). In recent years, more research has paid attention to the subjective qualities of places. For example, Chen~\etal~\cite{chen2023sensemap} developed SenseMap to analyze how a city is perceived in terms of qualities such as comfort and livability. Yang~\etal~\cite{yang2024emogeocity} proposed EmoGeoCity, a digital humanities system that analyzes and visualizes emotional information from the historical and literary materials of a city. A set of personal visualization systems have also been developed to help users recall places they have visited and cherished moments~\cite{thudt2015visual,carpendale2017subjectivity}.
However, to the best of our knowledge, there is still no systematic research on how geographic visualizations can be combined with affective design (\ie \textit{affective geovisualization design}). 

To understand this field more in-depth, we approach it from a two-phase focus: (i) incorporating the theoretical perspective of human-place relationships from geography to characterize affective geovisualization design features, and (ii) identifying different paradigms within this highly interdisciplinary field to draw insights from their practices.
Specifically, we manually collected a corpus of \ncorpus affective geovisualization designs and utilized the Person-Process-Place (PPP) model~\cite{scannell2010defining} to systematically code our corpus. As a result, we identified a rich array of methods for eliciting and communicating emotions, which help extend the existing taxonomy of affective visualization design~\cite{lan2023affective}. Next, the designs in the corpus were further clustered into four categories, each corresponding to a representative paradigm (\eg computational paradigm, anthropomorphic paradigm). These clusters not only help delineate the various interdisciplinary paradigms underlying the design practices but also offer high-level guidance for future practitioners.

In summary, our contributions include: (i) We present an incremental work for the research on affective visualization design. By focusing on a specific sub-area, namely affective geovisualization design, we contribute additional empirical examples and geo-spatial specificity to existing research. (ii) We incorporate the PPP model to help interpret affective geovisualization design, which enhances the theoretical depth and focused perspective of our research compared to traditional open coding. (iii) We present a rich array of low-level design methods and high-level design paradigms, offering practical insights for the creation of affective geovisualization design.
\section{Related Work}
\label{sec:related}

In this section, we review previous work about affective visualization design, emotional geography, and geographic visualization.

\subsection{Affective (Geo)visualization Design}
\label{ssec:affectivevis}

Affective visualization design uses data visualization to communicate or influence emotions~\cite{lan2023affective}. This thread of research is characterized by the acknowledgment of the presence and legitimacy of emotions in visualization creation and dissemination. For example, Lee~\etal~\cite{lee2022affective} argued that data visualization is never objective and proposed a taxonomy for the affective intents of designers (\eg strengthen a value or belief). Kennedy and Hill~\cite{kennedy2018feeling} found that people's emotions toward visualizations can arise from various elements, ranging from the data itself to the design style, subject matter, and source. Additionally, emotion has been found to be useful in enhancing user engagement and memory~\cite{lan2021kineticharts, gilmartin1991effects, boy2017showing}. When used appropriately, it can work hand in hand with rationality to enhance data comprehension and message absorption~\cite{anderson2021affective,lan2022negative}.
In terms of how to perform such design, factors such as color~\cite{bartram2017affective}, animation~\cite{lan2021kineticharts}, illustration~\cite{garreton2023attitudinal}, 
interaction~\cite{shi2022breaking,lan2022negative}, and physicalization~\cite{wang2019emotional} have been found effective in eliciting emotion.

When examining the nuances of affective visualization design, affective geovisualization design is a particularly noteworthy sub-type. According to the survey by Lan~\etal~\cite{lan2023affective}, geographers have not only shown an early interest in emotion but have also explicitly introduced concepts such as \textit{affective geovisualization}~\cite{aitken2011affective} and \textit{affective atlas}~\cite{cartwright2008developing}, driven by the rise of emotional geography (see \cref{ssec:emogeo}).
For example, in the 1990s, Gilmartin~\cite{gilmartin1991effects} studied the impact of map projection methods on users’ psychological perception of distance and emotional engagement. Nold~\cite{nold2009emotional} had participants wander around the city wearing sensors and visualized their feelings toward different places, thereby facilitating exploration and reflection on the urban environment.
Meanwhile, given the pervasiveness of geographic data in today’s digital world, some researchers in the visualization community have also explored affective geovisualization design in their systems and tools. For example, systems such as SenseMap~\cite{chen2023sensemap} and EmoGeoCity~\cite{yang2024emogeocity} have been developed to analyze how a city is perceived, adding an emotional layer to the conventional visualization of the physical environment. Thudt~\etal~\cite{thudt2015visual} proposed Visual Mementos, which visualizes GPS data to help individuals reminisce and share personal experiences. 

However, despite its early emergence and rich applications, to the best of our knowledge, little research has been done to systematically examine affective geovisualization design. By conducting a focused analysis of this area, this work aims to contribute an incremental work for the research on affective visualization design.

\subsection{Emotional Geography}
\label{ssec:emogeo}

In 2001, Anderson and Smith~\cite{anderson2001emotional} introduced the concept of \textit{emotional geographies}, emphasizing the key role of emotion in studying how humans interact with places and spaces. Since then, a series of conferences, journal issues, and books focused on emotional geography have emerged, establishing this area as a significant substream within geographical research~\cite{bondi2016emotional}.
According to the summarization by Bondi~\cite{bondi2017making}, emotional geography was grounded in a set of geographical traditions that were deeply influenced by philosophies and methodologies from the humanities and social sciences, such as humanistic geography, feminist geography, and non-representational geography. Rather than focusing on the scientific laws governing the Earth and its atmosphere (\eg climate, hydrology, soil), these research threads exhibit strong attention to human and their subjective experience with places.  
For example, \textit{humanistic geography} sought to understand the subjective and experienced ``life-world'', seeing places as a center of meaning constructed by personal experience~\cite{tuan1975place,tuan2017humanistic}.
Under this epistemology, researchers have found that the same location can evoke different ``senses of place'' among individuals, influenced by personal histories and socio-cultural contexts (\eg residential status, age stage)~\cite{hay1998sense,milligan2016healing}. Even the same geographic scale (\eg the distance between work and home) can be perceived differently given different moods and states (\eg the road feels lengthy and tiring after a day of overtime work, but becomes shorter and more pleasing on the last day before a holiday)~\cite{ley2014humanistic}. 
Sometimes, individuals may even exhibit extreme emotions to places, such as agoraphobia (\ie an intense fear of open or crowded spaces)~\cite{davidson2017phobic} and madness~\cite{parr1999delusional}.
The deconstruction of absolute objectivity and the embrace of subjective emotions in the field of geography highly resonate with the evolving perspectives in the visualization community (as introduced in \cref{ssec:affectivevis}). Meanwhile, the flourishing of this line of research has provided valuable perspectives for our work. For example, Scannell and Gifford~\cite{scannell2010defining} proposed the Person-Process-Place model to help analyze place attachment, and this framework has been widely applied to understand the emotional bonds between individuals and their environments. We believe that incorporating such theoretical resources from emotional geography will be highly beneficial for our analysis.


\subsection{Geographic Visualization}
\label{ssec:geovis}

Map is the earliest form of visualization~\cite{friendly2001milestones} and geographic visualization has always been a critical theme in visualization research. For example, abundant analytical work has been done to mine and explore geographic data~\cite{feng2022survey} (\eg trajectory data~\cite{chen2017vaud, he2019interactive, weng2020towards, andrienko2021visual}, point of interest data~\cite{chen2023sensemap, li2020warehouse}, geo-text data~\cite{cao2012whisper}, telco data~\cite{wu2017exploring}, image data such as google street view~\cite{shen2017streetvizor}, meteorological data~\cite{qu2007visual}).
Also, a series of studies have been done to optimize the information design and interaction methods for geographic visualizations. For example, LiberRoad~\cite{guo2023liberroad} employs hierarchical circles to represent spatial uncertainty and movement patterns in geospatial data. Nusrat~\etal~\cite{nusrat2016evaluating} compared the effectiveness of four cartogram designs and found that contiguous cartograms were the most effective overall while each showed strengths in specific tasks. 
Go{\l}biowska and {\c{C}}{\"o}ltekin~\cite{golbiowska2020rainbow} investigated the effectiveness of rainbow color schemes on maps. Roth~\etal~\cite{roth2013empirically} developed a taxonomy of interaction primitives for map-based visualizations, including objectives (\eg identification, comparison) and operators (\eg filtering, zooming).

On the other hand, some studies have focused on the communicative aspects of geographic visualizations. For example, Wood~\cite{wood1992power} and Harley~\cite{harley2008maps} pointed out that maps can carry inherent subjectivity, such as the perspectives of cartographers and the will of those in power; maps also have an extensive history of being used for propaganda and misleading viewers. Roth~\cite{roth2021cartographic} discussed how to integrate narrative elements with map design to create persuasive and emotionally resonant visualizations.
Muehlenhaus~\cite{muehlenhaus2012if} compared four rhetorical styles (\ie authoritative, understated, propagandist, and sensationalist) in map design.
In recent years, geographic visualization has also been frequently observed in data storytelling.
For instance, 
Li~\etal~\cite{li2023geocamera} examined camera movement strategies in geographic data videos. 
Latif~\etal~\cite{latif2021deeper} explored the interplay between visualizations and text in geographic data-driven stories, analyzing how these elements reference each other and interact to enhance comprehension and engagement.
Slavik~\etal~\cite{slavik2024using} discussed how geovisualizations can be used to educate the public about environmental health hazards. 

Similar to the above research, this work examines human factors in geographic visualization and specifically focuses on how geographic visualization can be utilized to communicate and influence emotion.

\section{Methodologies}
In this section, we describe how we constructed our corpus, refined our research scope, and coded the corpus.

\subsection{Scoping Review}

Affective geovisualization design is highly interdisciplinary, and to the best of our knowledge, no mature corpus specifically targeting this area has been established. Therefore, we decided to conduct a survey of relevant literature and identify qualified design works both from the papers themselves and from the in-the-wild projects they cited, following prior methodology~\cite{lan2023affective}.
We initiated our literature search using Google Scholar as the primary platform, using keywords such as \textit{emotional geography}, \textit{emotional cartography}, and \textit{affective geovisualization}. Papers that contain the required keywords in titles or abstracts were carefully reviewed. This step yielded an initial set of 227 papers. 
When we read through these papers, we noticed that there exist broader terms and concepts commonly used by scholars from various interdisciplinary fields. This situation is very similar to some previous studies (\eg \cite{fu2023more}, \cite{long2020ai}), which have shown that a single round of keyword searching is insufficient to cover papers from different disciplines. Therefore, we also adopted the scoping review method~\cite{arksey2005scoping} used in these studies, which involves iterative keyword identification and complementary searching to ``map rapidly the key concepts underpinning a research area and the main sources and types of evidence available... especially where an area is complex or has not been reviewed comprehensively before''.

Guided by this, we marked additional keywords including \textit{personal geovisualization}, \textit{psychogeography}, \textit{humanistic geography}, \textit{phenomenological geography}, and \textit{feminist geography} from the first round of literature review and conducted a second-round search, which led to 71 additional papers.
The third-round search complemented the corpus with papers about \textit{affective visualization design} and \textit{geographic data storytelling} (\eg \cite{lan2023affective,mayer2023characterization,latif2021deeper,li2023geocamera}). These papers were mainly published in the visualization community and may not directly mention concepts such as \textit{emotional geography}, but they have the potential to contain affective geovisualization designs in the empirical materials presented or cited.
After three rounds of search, we collected a total of 313 papers.
We also documented the primary disciplines tagged to these papers on Web of Science. For papers not indexed in Web of Science, we manually assigned disciplines based on the venues' official descriptions~\cite{lan2023affective}. The papers span diverse disciplines and venues, such as geography (e.g., The Cartographic Journal), computer science (e.g., IEEE Transactions on Visualization and Computer Graphics), social science (e.g., Borders in Globalization Review), and psychology (e.g., Visual Methods in Psychology).

\subsection{Identifying Design Works}
\label{ssec:identify_design}

\begin{figure}[h]
    \centering
    \includegraphics[width=1\linewidth]{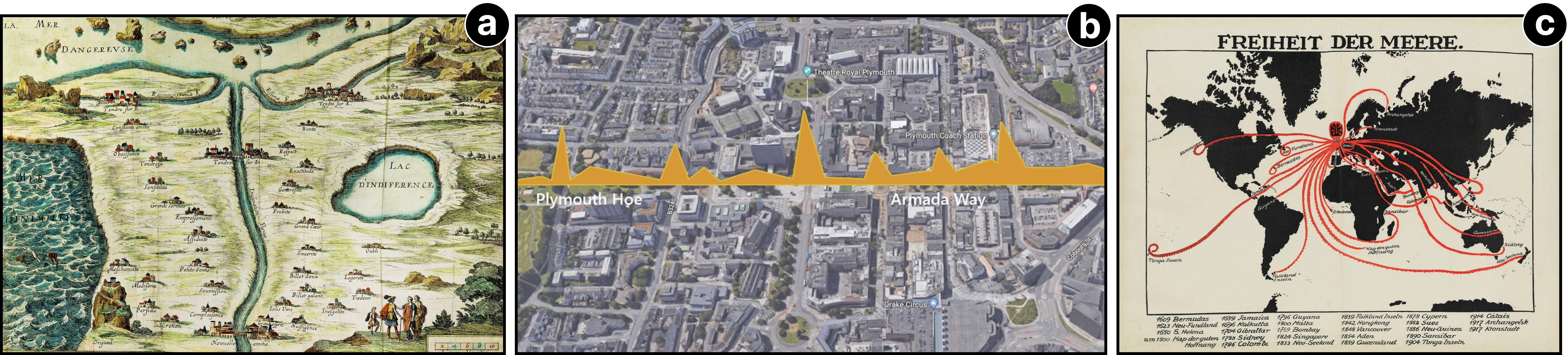}
    \caption{Counterexamples. (a) Map of Tendre~\cite{bruno2002atlas}. (b) Average stress levels measured by Galvanic skin response (GSR) of all participants mapped to Plymouth City Centre, showing peaks of stress at crossings and junctions~\cite{fathullah2018engaging}. (c) A poster depicts Britain as an octopus threatening the "Freiheit der Meere," Freedom of the Seas~\cite{wood1992power}.}
    \label{fig:excluded}
\end{figure}

Next, we attempted to identify affective geovisualization designs from the pool. We read all of these papers and manually filtered them according to the following criteria: (i) The paper must include geographic data visualization(s). This criterion led to the exclusion of papers that only introduced the theories of emotional geography (\eg \cite{anderson2001emotional,davidson2004embodying}) and those that studied how people respond emotionally to places but did not incorporate any visual methods (\eg \cite{milligan2016healing, davidson2017phobic}). (ii) If geovisualization does exist, it should be geographically data-driven. This criterion led to the exclusion of works that resemble data visualizations in appearance but do not encode any real data. For example, the \textit{Map of Tendre} (\cref{fig:excluded} a) is an allegorical map based on the theme of love and depicts an imaginary land~\cite{bruno2002atlas}. 
It includes places like Villages of Sensibility, Sincerity, and Generosity, as well as perilous regions such as the Lake of Indifference. These elements symbolize the various stages of love.
(iii) Following the definition of affective visualization design~\cite{lan2023affective}, we required that there should be explicit descriptions indicating that the geovisualization was intentionally designed to communicate or influence emotion. Thus, for example, data visualizations used purely for analysis were excluded (\eg \cref{fig:excluded} b).
(iv) The visualization should be grounded in empirical spatial experiences. This criterion excluded solely rhetorical geovisualizations such as propaganda maps  (\eg \cref{fig:excluded} c), where geographic elements function as rhetorical constructs rather than spatial realities, and emotional encodings originate from ideological agendas instead of human activities. While such maps technically satisfy the definition of affective visualization design (\ie intentional emotional communication), their analytical frameworks primarily reside in rhetorical and political studies and may bring about the issue of paradigm incommensurability.

Therefore, to enhance research feasibility and focus, and to avoid issues related to conflicting theoretical frameworks, we ultimately retained 62 designs. All of these designs contain real geographic data, embody authentic people-place relationships, and are affective. 
All of these designs, along with their associated materials (papers, online sources, and authors' personal websites), can be accessed at \url{https://affectivegeovis.github.io}.

\subsection{Coding Process}

\begin{figure}
    \centering
    \includegraphics[width=0.9\linewidth]{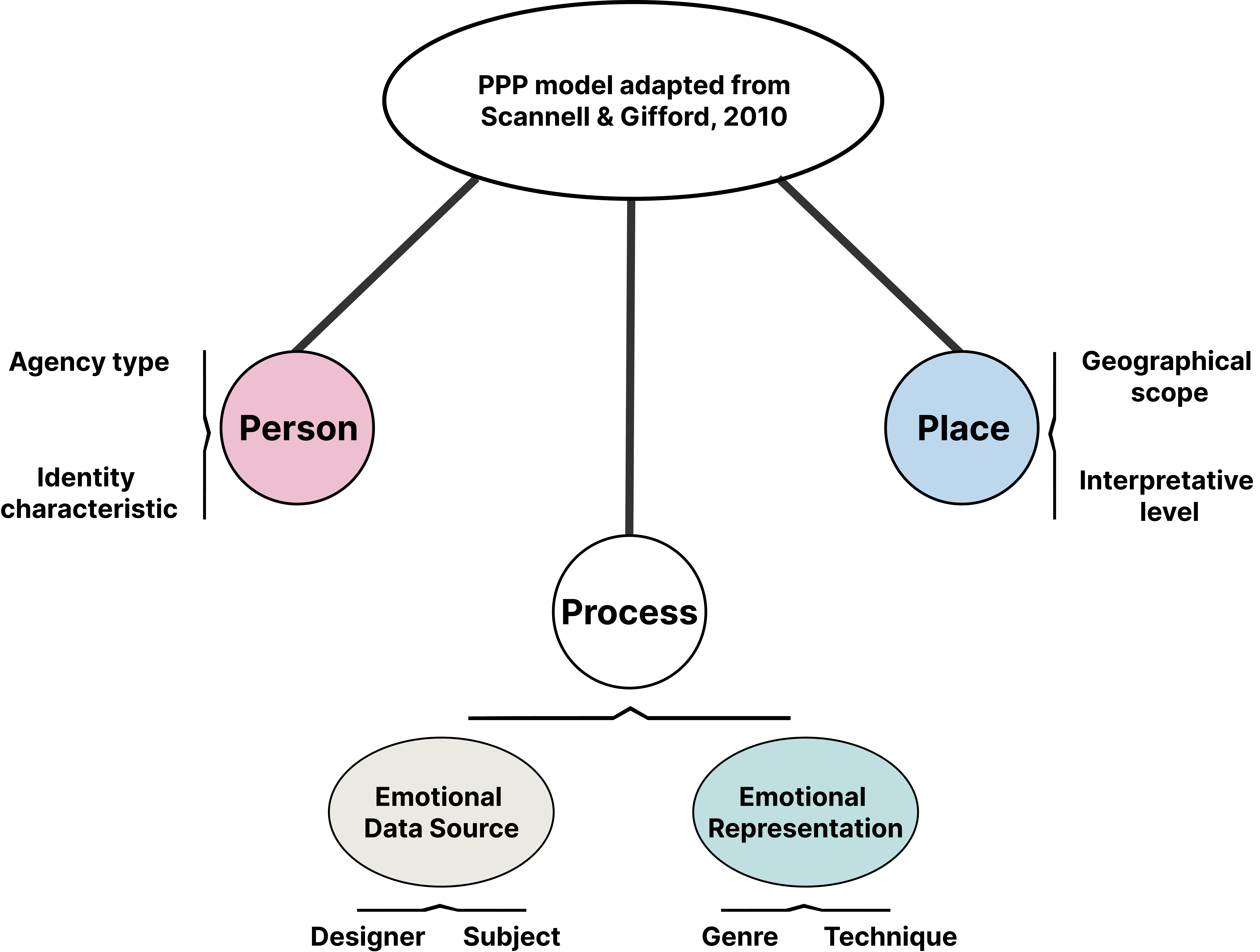}
    \caption{The Person-Process-Place model.}
    \label{fig:ppp}
    \vspace{-2em}
\end{figure}


\begin{sidewaystable*}
    \centering
    \includegraphics[width=\textheight]{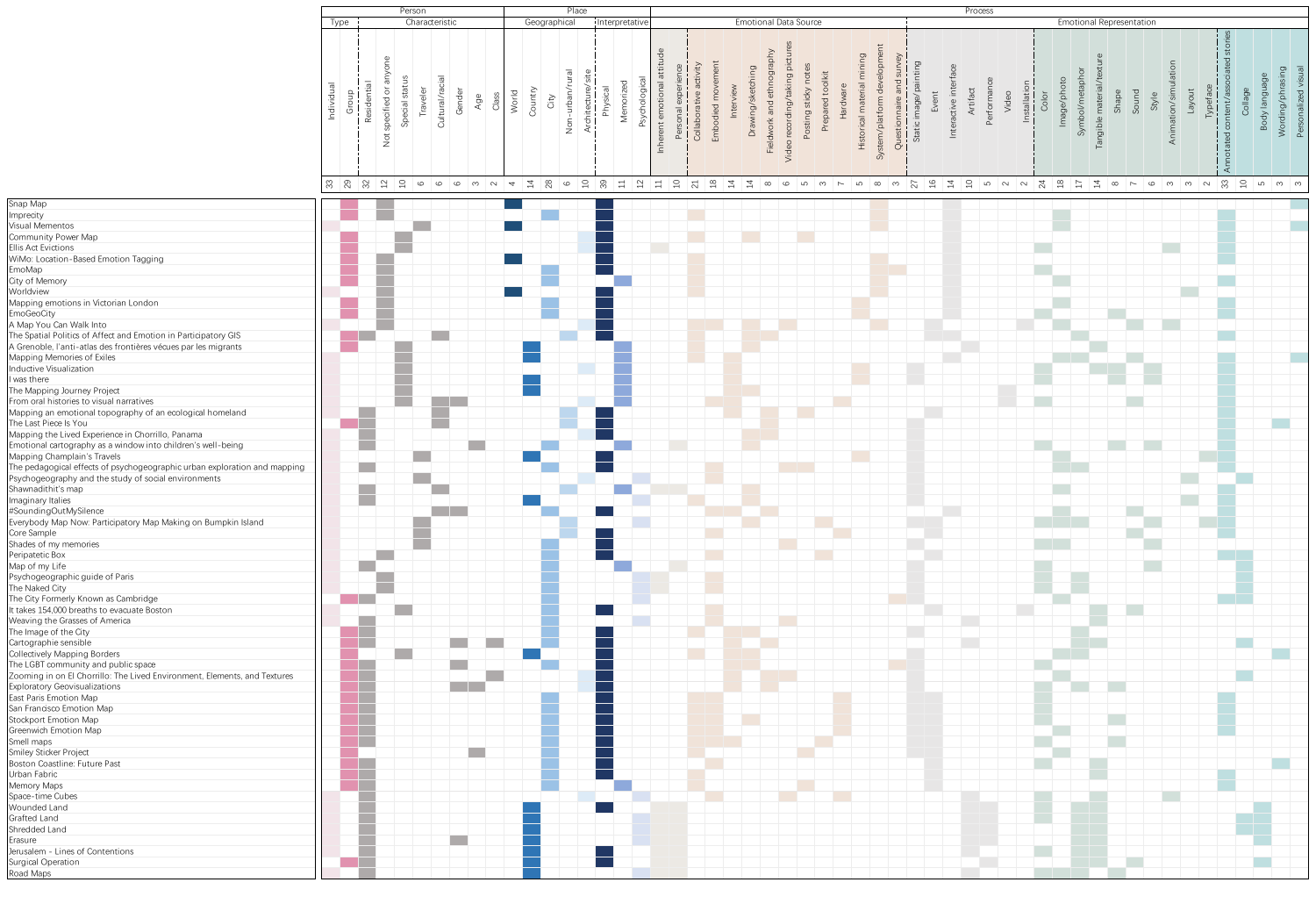}%
    \caption{A taxonomy of affective geovisualization design. The rows correspond to all the design works in our corpus, while the columns correspond to the coded dimensions adapted from the PPP model. An interactive version is available at \url{https://affectivegeovis.github.io}.}
    \label{tab:designspace}
\end{sidewaystable*}

Based on the corpus, we referred to the Person-Process-Place (PPP) model~\cite{scannell2010defining} in humanistic geography to guide our coding process. The PPP model is a classic research framework proposed to help explain people-place attachment and consists of three main dimensions: (i) Person, namely who is experiencing the connection with places, such as whether it is an individual or a group; (ii) Place, the traits of geographic space, such as whether it is a physical building or a social arena; and (iii) Process, which describes how the person forms a connection with the place. In the original model, the process dimension is quite broad and can be divided into affective, cognitive, or behavioral processes.
We believe this framework is highly compatible with our research, as it has been frequently utilized to explain subjective people-place relationships~\cite{counted2016making, al2025unearthing}. Additionally, since it originated in geography research, it naturally pays attention to geographic attributes such as the characteristics of places. Compared to general coding, drawing on an existing research framework can help us gain more in-depth insights specific to geographic context.

We adapted the PPP model to our specific research questions and developed the following coding scheme~(\cref{fig:ppp}): (i) For the \person{\textbf{person}} dimension, we coded two sub-dimensions, including the agency types (\eg individual, group) and their identity characteristics (\eg cultural, gender); (ii) For the \place{\textbf{place}} dimension, we coded two sub-dimensions, including geographical scopes (\eg city, country) and interpretive levels (\eg realistic, memorized); (iii) For the process dimension, in the original model, it can refer to any psychological and physiological interactions between people and places. However, in our study, we are interested in how affective geovisualization design acts as a specific mediating process to construct or empower people-place relationships. Therefore, we operationalized this dimension into two sub-dimensions closely related to data visualization: \processi{\textbf{emotional data sources}} and \processii{\textbf{emotional representations}}. Below is an coded example: \textit{East Paris Emotion Map} project (\cref{fig:examples} B) invited a group of citizens (\person{\textit{group, residential}}) to wander (\processi{\textit{embodied movement}}) in Paris (\place{\textit{city, physical place}}) while wearing bio mapping sensors (\processi{\textit{hardware}}). Their emotional data was collected during walking and then reflected in a workshop (\processi{\textit{collaborative activity}}). The resulting map synthesized the participants' collective emotional arousal using red (\processii{\textit{color}}), associated with participants' annotations on the map (\processii{\textit{annotated content}}). 
Detailed explanations of codes are discussed in ~\cref{sec:results}.

Two authors conducted {deductive coding~\cite{elo2008qualitative} on our corpus based on the codebook introduced above with a top-down approach. 
One coder has a humanities background with expertise in visual communication and affective design, while the other has a technical background with expertise in developing geo-spatial visual analytics systems. This expertise is both highly relevant to the research topic, while also providing complementary interdisciplinary knowledge and perspectives.
Each coder first reviewed the materials and coded the corpus independently based on the codebook, and wrote down notes with supporting evidence. Subsequently, we discussed all disagreements and marked questions through three iterative meeting. During the meetings, we focused on discussing disagreements and marked uncertainties, and then revised the codebook when reaching consensus. Also, the wording of each code was iteratively adjusted to ensure clarity and conciseness. 
For example, our initial codebook contained a code called \textit{workshop}, which is a term frequently found in the designs' associated descriptions. However, during our coding process, we found that the meaning of a \textit{workshop} can be quite vague and broad. For instance, a workshop might include a variety of activities such as drawing, sticking Post-it notes, and interviews. Therefore, we ultimately decided to remove the code \textit{workshop} and replace it with more specific activity forms, such as \textit{drawing}, \textit{posting sticky notes}, and \textit{interview}.
As another example, we initially used the code, \textit{dérive/walk}, to describe practices related to urban drifting, as it was a classic concept posed by Guy Debord~\cite{debord1958theory} in the 20th century and was cited by many designers. However, during the coding process, we noticed that this method, which originally emphasized walking, had evolved to include new forms such as driving or running. Therefore, we finally used \textit{embodied movement} to encompass these movement-related activities.

\section{A Taxonomy of Affective Geovisualization Design}
\label{sec:results}
The results are illustrated in \cref{tab:designspace}. Below, we present the findings of the three main dimensions and their sub-dimensions one by one.

\subsection{Person}
\label{ssec:person}
This dimension describes \person{\textbf{who}} has emotional connection with places.

\textbf{Agency types.} 
People, as active agents, can form affective connections with geographic spaces as individuals or in groups. In our corpus, we identified 33 cases where emotions originated from \person{\textit{individuals}}. For example, in \cref{fig:examples} A, the designer documented her movement data and the organic materials that touched her emotionally during a walk, and then froze them into ice cubes~\cite{gardener2017walk, gardener2018interdisciplinary}. This data physicalization reflects her private subjective experience.
By contrast, we also identified 29 cases where emotions originated from \person{\textit{groups}}. For example, the \textit{East Paris Emotion Map} project (\cref{fig:examples} B) invited a group of volunteers to walk around Paris together while wearing sensors, and visualized their emotional data collectively on a map~\cite{nold2009emotional, nold2018bio}.

\begin{figure*}[t]
    \centering
    \includegraphics[width=0.9\linewidth]{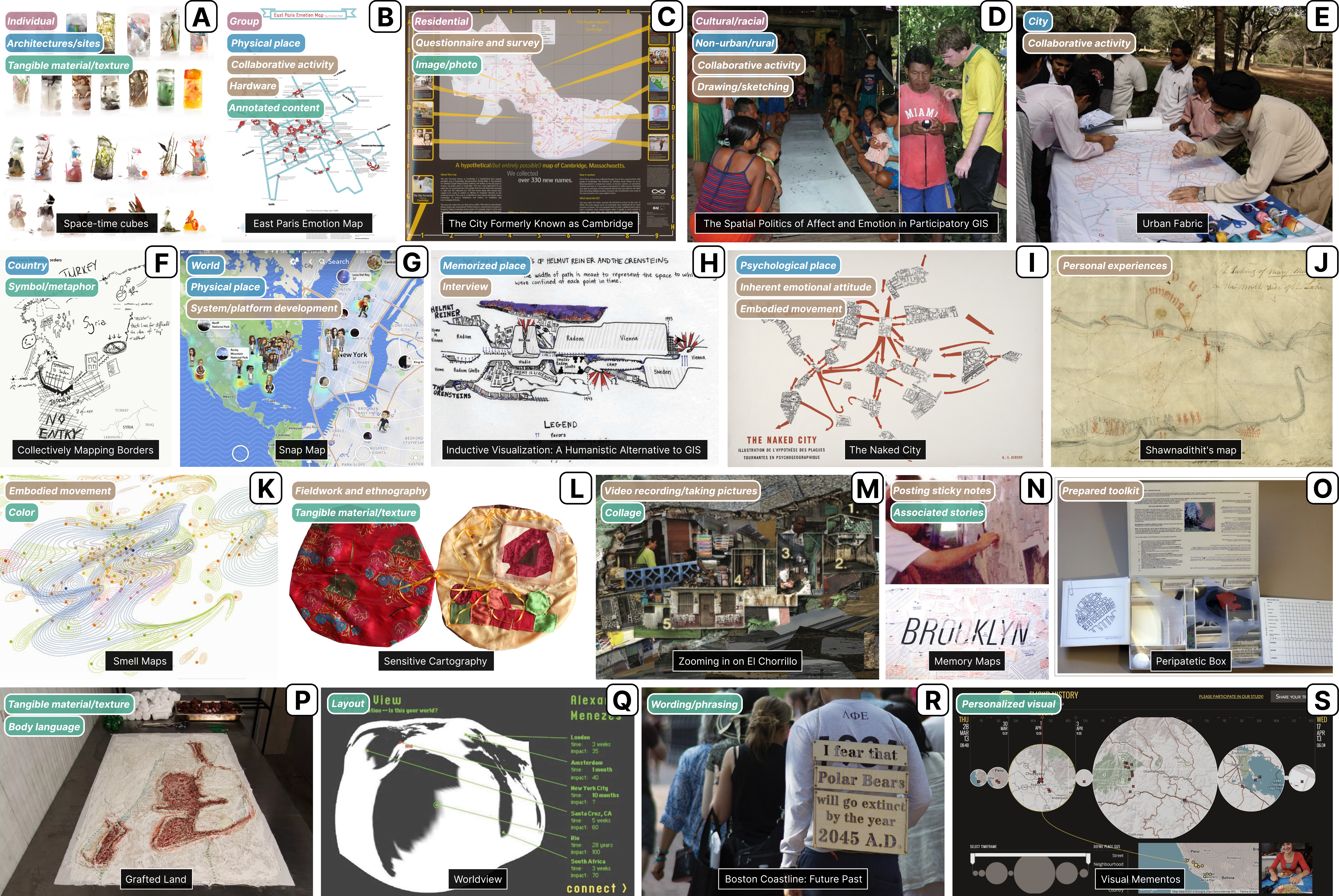}
    \caption{Examples of affective geovisualization designs. Each work is annotated with the corresponding codes referenced in the text.}
    \label{fig:examples}
\end{figure*}

\textbf{Identity characteristics.}
Although much research in the visualization community often labels people with abstract descriptions such as ``users'', ``general public'', and ``laypeople''~\cite{burns2023we}, we noticed that the vast majority of design projects (80.6\%) in our corpus have involved people with specific, concrete identities, rather than treating them as generic, decontextualized data points.
For example, we identified 32 cases that focus on the \person{\textit{residential}} identities of people, with the most typical being the attention to people's feelings toward their living spaces, such as their neighborhoods, cherished places, and familiar routes.
As an example, \textit{The City Formerly Known as Cambridge} (\cref{fig:examples} C) invited people to rename the sites in Cambridge based on their personal understanding and memories.
We also identified 10 cases that pay attention to people in \person{\textit{special statuses}}, especially those who face political and legal challenges, such as refugees.  \textit{The Mapping Journey Project}, for example, invited eight individuals, who were forced into illegal travel due to political circumstances, to trace their routes on maps using markers and narrate the stories of their secret journeys~\cite{moro2021mapping}.
Additionally, 6 cases focus on \person{\textit{travelers}}, who form emotional connections with places during short-term tourism or visits. Six cases are centered on \person{\textit{cultural/racial}} (\eg Muslim) and \person{\textit{gender}} (\eg women) identities, respectively. Three cases are about \person{\textit{age}} identities (\eg the elderly and children), and two are about \person{\textit{class}} identities (\eg the poor).
For example, \cref{fig:examples} D organized a participatory event where an indigenous community in Peru collaboratively created a map of their land, marking cultural heritages with symbols~\cite{young2013spatial}.
Twelve cases (19.4\%) do not describe the specific identities of the people involved. 

\subsection{Place}
\label{ssec:place}

This dimension refers to \place{\textbf{where}} emotional connections are situated.


\textbf{Geographical scopes.}
The designs in our corpus deal with varying geographical scopes. Among these, \place{\textit{city}} (28) is the most common. As the dwellings with the densest populations on Earth, cities embody a wide range of feelings and are particularly favored by designers for exploration. 
For instance, \textit{Urban Fabric} (\cref{fig:examples} E)~\cite{iturrioz2010artistic} invited participants to stitch symbols onto a hand-embroidered city map, marking places that held positive and negative significance for them. 
Another commonly addressed scope is \place{\textit{country}} (14). \cref{fig:examples} F, for example, shows a collective effort to redraw Syria's borders based on interviews with refugees, investigating their subjective experiences towards the country and its bordering areas~\cite{kelly2016collectively}.
Ten cases focus on emotional connections with specific \place{\textit{architectures/sites}}, such as \cref{fig:examples} A, which reflect the designer's sensory experience in a park~\cite{gardener2017walk, gardener2018interdisciplinary}.
Six cases focus on \place{\textit{non-urban/rural places}}, such as islands and jungles (\eg \cref{fig:examples} D).
Lastly, 4 cases visualize emotions at the \place{\textit{world}} level, such as \textit{Snap Map} (\cref{fig:examples} G)~\cite{roth2021cartographic}, a globally accessible map platform that allows users to share their GPS locations and emotional states.

\textbf{Interpretative levels.} 
Another interesting finding from our work is that places are subject to human interpretation, thus exhibiting varying degrees of subjectivity. 
In our corpus, we identified three levels of interpretativity related to places. 



\place{\textit{Level 1: Physical place}} (39).
Physical places encompass embodied experiences rooted in physical co-presence, ranging from immediate affective responses to accumulated affective imprints from long-term inhabitation. Key to this category is the corporeal accountability of real places, which means these experiences remain tied to verifiable geographic coordinates and habitual movement patterns. Their corresponding visualizations typically include clear information such as base maps, topological information, and place names.
For example, applications such as \textit{Snap Map} (\cref{fig:examples} G)~\cite{roth2021cartographic} allow users to share their locations and associated emotions based on GPS. Offline activities, such as \cref{fig:examples} B, enable participants to engage with specific places, experiencing real-time emotions in physical spaces~\cite{nold2009emotional, nold2018bio}.

\place{\textit{Level 2: Memorized place}} (11).
Memorized places emerge through reconstruction of geographic information recalled by the involved subjects. Due to memory decay or reprocessing, the information undergoes cognitive compression, emotional amplification, or chronological scrambling. As a result, the depiction can be vague, imprecise, or selective. Therefore, its level of interpretability is higher than that of level 1, and the visualizations typically include scattered and collage geographic information, as well as a significant amount of missing values or uncertainty data.
For example, \cref{fig:examples} H is a map that visualizes the personal experiences of Holocaust survivors based on their testimonies. These memories are closely tied to subjectively experienced, emotionally significant places, many of which are unprojectable due to uncertain locations or misheard place names~\cite{knowles2015inductive}.

\place{\textit{Level 3: Psychological place}} (12).
At the most subjective extreme, psychological places are more likely to be unbound from physical laws or historical accuracy. Although there still exists real geographic data (as required by us in \cref{ssec:identify_design}), such data may play a minor role or serve more subjective purposes. Correspondingly, their visualizations often contain symbolic expressions, such as metaphorical geometries or speculative interfaces that deviate from scientific spatial logic.
For example, \cref{fig:examples} I is a psychogeographic map created by Guy Debord in 1957. This work selected 19 separate landmarks in the city of Paris. While retaining the appearance of these landmarks and their relative geographical positions, they were collaged and connected with red arrows. This design served Guy's political and artistic views (\ie Situationism) at the time—he saw cities moving towards industrialization, spectacle, and homogenization under the influence of capitalism. Therefore, he advocated exploring cities through aimless dérive (drifting) to challenge the conventional understanding of urban space and to discover subjectively meaningful urban experiences~\cite{pinder1996subverting, moro2021mapping}.

\subsection{Process}
\label{ssec:process}

This dimension describes \textbf{how} visualization design can be used to represent or enhance the emotions arising from person-place relationships. It primarily functions in two key facets: the emotional content of the data, and the emotional expression in the outer presentation.

\subsubsection{Emotional data sources}

Regarding \processi{\textbf{how}} emotion can be infused into data, we observed that it can either originate from the designers themselves or from others.

\textbf{Originate from designers.}
We identified 11 works whose emotion stems from the designers' \processi{\textit{inherent emotional attitude}} (\eg skepticism, rebellion, negation), as seen in \cref{fig:examples} I, which we discussed earlier.
We also identified 10 works where emotion emerges from designers' \processi{\textit{personal experiences}}. 
For example, \textit{Shawnadithit's map} (\cref{fig:examples} J) was drawn by the last member of the Beothuk, a group of indigenous people of Canada. The map conveys the pain and bodily interaction with space, documenting the Beothuk's experience of facing oppression and conflict in their homeland~\cite{nast1998places}.

\textbf{Originate from other people.}
Interestingly, in our corpus, a greater number of works are not expressions of the designers' own emotions, but rather grant the right to express emotions to others. Meanwhile, we observed that very diverse methods have been employed to evoke participants' emotions towards places. 




\begin{table}[h]
    \centering
    \fontsize{7pt}{8pt}\selectfont
    \begin{tabular}{c p{0.45\linewidth} c}
    \hline
        \textbf{Category} & \textbf{Method} & \textbf{Count} \\ 
        \hline
        Qualitative & Collaborative activity & 21 \\ 
         & Embodied movement & 18 \\
         & Interview & 14 \\
         & Drawing/sketching & 14 \\
         & Fieldwork and ethnography & 8 \\
         & Video recording/taking pictures & 6 \\
         & Posting sticky notes & 5 \\
         & Prepared toolkit & 3 \\
        \hline
        Quantitative & Hardware & 7 \\ 
         & Historical material mining & 5 \\
        \hline
        Mixed & System/platform development & 8 \\ 
         & Questionnaire and survey & 3 \\ 
        \hline
    \end{tabular}
    \caption{Methods that evoke participants' emotional connections.}
    \label{tab:process1}
\end{table}

We have summarized these identified methods in \cref{tab:process1}. Among them, most methods are qualitative:

\processi{\textit{Collaborative activity}} (21). The most frequently used approach is to organize activities that involve social interaction. In this process, people are given the chance to experience or express their emotions by actively participating in events such as workshops and co-design. For example, in \cref{fig:examples} B, D and E, participants were invited to explore places or express emotions toward places through collective activities such as reminiscing or creating. These activities embedded emotional content and personal narratives into the visualization.

\processi{\textit{Embodied movement}} (18). As introduced earlier, concepts such as dérive proposed by Guy Debord encourage people to explore spaces and places with their own bodies~\cite{debord1958theory}, and his \textit{The Naked City} (\cref{fig:examples} I) is an early psychogeographic map that exemplifies this method.
To date, this method has been widely adopted, as seen in various visualization designs that employ urban drifting, such as \cref{fig:examples} K, where participants wander through the city to perceive and document scents~\cite{mclean2019nose, mclean2020temporalities}.


\processi{\textit{Interview}} (14). Interviews are a common method for starting conversations with others and understanding interviewees' emotional experiences and opinions. For example, in \cref{fig:examples} H, interviews were conducted with Holocaust survivors. These personal narratives vividly recount a wide range of experiences and emotions anchored in specific places and times, such as terrifying near-death experiences~\cite{knowles2015inductive}.

\processi{\textit{Drawing/sketching}} (14). The act of drawing and sketching is an effective method and is often used in psychology for emotional expression and self-mining.
For example, in \cref{fig:examples} D, by drawing the map of their land and culture, the locals fostered a sense of pride~\cite{young2013spatial}.

\processi{\textit{Fieldwork and ethnography}} (8). Some researcher approach the emotion of subjects by getting involved in their lives. For example, to create \textit{Sensitive Cartography} (\cref{fig:examples} L), the designer spent several months living with the women of a disadvantaged neighborhood in Marrakech, following them in their daily lives, both in their domestic spaces and as they moved through the city.
Through this approach, the feelings and perceptions of the studied women were able to be sensed by the author and expressed through creation.

\processi{\textit{Video recording/taking pictures}} (6). This method often involves using smartphones or cameras to capture content that holds emotional significance. 
For example, in \cref{fig:examples} M, the author used photography to document the textures and architectural details of El Chorrillo, Panama City, capturing resident’s perspectives of spaces, as well as her own experience with encountering the neighborhood~\cite{powell2010making}.

\processi{\textit{Posting sticky notes}} (5). This method enables participants to externalize their emotions through writing on or posting sticky notes.
For example, \textit{Memory Maps} (\cref{fig:examples} N) invited citizens to share their stories of the city. They were asked to write these stories down and then pin them to a map~\cite{krygier2006jake}.

\processi{\textit{Prepared toolkit}} (3). Some projects provided participants with prepared toolkits, which often include cards, stickers, or papers, to facilitate creative expression while exploring places and constructing maps. In \cref{fig:examples} O, users were given a \textit{Peripatetic Box} to document their journeys facilitated by writing on cards, drawing maps, and collecting materials. This process allows them to add a personal, emotional layer to their experiences of cities~\cite{moro2012peripatetic}.

Two methods are quantitative:

\processi{\textit{Hardware}} (7). 
Devices such as biosensors can be used to capture, record or elicit emotional responses across spatial contexts. For example, \cref{fig:examples} B used Bio Mapping—a technique integrating biometric sensors (e.g., galvanic skin response monitors) with geolocation tracking—to quantify participants' affective states as they navigate urban environments~\cite{nold2009emotional, nold2018bio}. 

\processi{\textit{Historical material mining}} (5). This approach extracts emotion from historical data, such as poems and fictions. For example, Pearce and Hermann mapped the emotional expressions from Champlain’s travel diaries, allowing readers to engage with his journey not just as a spatial record but as an emotional and experiential account~\cite{pearce2010mapping}.

Besides, we also identified two mixed methods.

\processi{\textit{System/platform development}} (8). This method is more function-oriented, enabling users to share emotions and experiences in real time by combining quantitative data processing (\eg GPS) with qualitative inputs (\eg comments). For example, \textit{Snap Map} (\cref{fig:examples} G) provides a public visualization where users can dynamically express their locations and the feelings related.

\processi{\textit{Questionnaire and survey}} (3). 
This method can be used to collect various types of data, ranging from quantitative ratings to qualitative insights, such as feelings.
For example, \cref{fig:examples} C gathered over 300 new names for public spaces in Cambridge, reflecting contributors' personal experiences and subjective perceptions.






\subsubsection{Emotional representations}



Regarding \processii{\textbf{how}} emotions can be expressed and presented in visualizations, we identified a total of 7 high-level genres and 15 low-level design techniques that were explicitly mentioned by their designers as affective. These genres are mostly consistent with those identified in general affective visualization design~\cite{lan2023affective}, with one newly discovered genre called \textit{performance} (typically live performances by artists interacting with the map). The 15 techniques were categorized into two strategies: sensations and narratives.

Firstly, emotions can be encoded into various sensory experiences:

\processii{\textit{Color}} (24) is a frequently adopted affective design technique. For example, in \textit{Smell maps} (\cref{fig:examples} K), colors were diversely used to depict the feelings of smells, such as ``dirty, even grimy'' colors reflecting ``Cheap food and coffee; CO/CO2 in the air; Parfum, especially near entrance'', or vibrant color reflecting ``greenery during winter''~\cite{mclean2019nose, mclean2020temporalities}. 
Next is the usage of realistic \processii{\textit{images/photos}} (18), and \processii{\textit{symbol/metaphor}} (17). Both techniques use figurative language to express emotion, with the main difference being that the former is realistic, while the latter is more implicit. 
For example, in \cref{fig:examples} C, a set of photographs contributed by participants were displayed on the map, visually narrating the emotional stories behind them.
\cref{fig:examples} F uses symbolic representations such as emojis and arrows to illustrate the complexities of border crossings, offering an emotional view of the refugee experience through metaphoric symbols~\cite{kelly2016collectively}.
Besides, the use of \processii{\textit{tangible materials}} (14), or physicalization, is also common. A wide range of materials have been chosen to express emotion, such as organics (\cref{fig:examples} A) and cloth (\cref{fig:examples} L).
\textit{Grafted Land} (\cref{fig:examples} P) even uses gauze and stitching to cover disputed areas on a Palestinian map, with the gauze appearing as if it is seeping blood, symbolizing the process of healing~\cite{littman2023mapping}.
Some designs used deliberate \processii{\textit{layout}} (3) to express emotion. For example, \textit{Worldview} (\cref{fig:examples} Q) remaps the world from the user's emotional perspective by first marking locations of personal importance on a world map and then distorting it through a ``fish-eye'' algorithm to fit the user's view~\cite{krygier2006jake}.
Other affective design techniques, including \processii{\textit{shape}} (8), \processii{\textit{sound}} (7), \processii{\textit{style}} (6), \processii{\textit{animation/simulation}} (3) and \processii{\textit{typeface}} (2), have also been used to communicate or influence emotion.


Additionally, five narrative methods that serve affective expressions have been identified:

Adding \processii{\textit{annotated content/associated stories}} (33) accompanied with visualization is the most common. 
For example, in \cref{fig:examples} B, after participants walked around the city wearing sensors, they were gathered to reflect on anecdotes or momentary sensations they experienced in the space, particularly in areas where the emotion value seemed to fluctuate significantly. Their narratives were recorded by the author and incorporated into the final map~\cite{nold2009emotional, nold2018bio}.
Or, designers may use the rhetoric of \processii{\textit{collage}} (10), which combines different elements to create open-ended stories, challenging established assumptions and helping to rebuild geographic narratives~\cite{powell2010making}.
As in \cref{fig:examples} M, the author employs collage to convey her disoriented experience of encountering El Chorrillo and to draw viewers into a closer observation of the reflective maps~\cite{powell2010making}.
Five cases use \processii{\textit{body language}} to tell the story. For example, in \cref{fig:examples} P, the artist, after wrapping the war-torn places with gauze, performed a surgical-style act to stitch the map, hoping it would ``heal''~\cite{littman2023mapping}.
We also identified 3 cases that use concise yet powerful \processii{\textit{wording/phrasing}} to influence emotion. For example, in the event \textit{Boston Coastline: Future Past} (\cref{fig:examples} R), participants held wooden boards displaying their own succinct emotional expressions. 
In addition, \processii{\textit{personalized visual}} (3) is also adopted as a technique to enhance emotional engagement by customizing the visualization to individual users. \cref{fig:examples} S, for instance uses digitalized personal data to create \textit{visual mementos}, enabling users to reflect memories and emotions by tailoring the data representation to their unique perspectives and experiences~\cite{thudt2015visual}.


\subsection{Difference with General Affective Visualization Design}

As an incremental work for the research on affective visualization design~\cite{lan2023affective}, this work identifies some consistent findings with prior work, particularly low-level design techniques such as using colors, typefaces, sound, and wording to communicate emotion. However, by focusing on affective geovisualization design specifically, we also observed three main distinctions:

\begin{itemize}[leftmargin=*]
    \item \textit{Fine-grained dimensions about the contexts of visualization.}
    Previously, the general framework~\cite{lan2023affective} does not dive deep into the contexts of emotional experience.
    By contrast, due to the importance of real-world environment and people-place relationships in geographic visualization, this study more closely examines \person{\textbf{who}} emotionally engages with the visualization and \place{\textbf{where}} it happens.
    As specific individuals come into analysis, they are no longer confined to being users or public but assume more agency types and identity characteristics. Similarly, places evolve beyond being mere spatial backdrops and emerge as lived, situated, and meaningful environments.
    Such a perspective enables a deeper exploration of how human experiences and spatial contexts co-construct emotional meaning and shape affective outcomes.
    \item \textit{Highlighting the infusion of emotion in data.}
    In affective geovisualization design, we found that not only is the visual representation crucial, but so is the data content. A diverse range of methods has been identified to help evoke or capture emotional data. These methods include not only quantitative, qualitative, and mixed forms, but also various innovative approaches, such as urban drifting and collaborative activities.
    This demonstrates that the affective visualization does not exist independently but relies on specific emotional semantics and contexts for people. Guided by these findings, the identified design methods have been refined into data-focused and visualization-focused categories in our taxonomy. 
    \item \textit{The prominence of metaphorical elements in affective geovisualization design.}
    In our analysis, we observed that a significant proportion (27.4\%) of examples explicitly use symbol or metaphor, which is more pronounced compared to the general affective visualization design~\cite{lan2023affective}. This may be due to the unique characteristics of geospatial data, which often carries deep cultural, historical, and personal significance. This inherent richness has led to its use in symbolic representations since early times. Therefore, as previous literature has noted, maps are inherently rhetorical and symbolic~\cite{harley1989deconstructing}. They do not merely depict physical spaces but also reflect the rich lived experiences of people and their interactions with the environment.
\end{itemize}

In a word, by contextualizing affective visualization design within emotional geographic practices, this taxonomy better supports the exploration of people-place relationship elicited, empowered, intensified, and represented by geographic data visualization.

\section{Pattern identification and observation}
\label{sec:patterns}
Although we have derived a relatively systematic taxonomy for affective geovisualization design, we feel it is also necessary to conduct some clustering and convergence of these codes to identify different paradigmatic threads in this interdisciplinary field, as well as to provide clearer entry avenues for researchers and practitioners interested in this area.
Specifically, in this section, by combining quantitative clustering with manual interpretation and verification, we identified four distinctive paradigms in affective geovisualization design.

\subsection{Method}
We adopted computational grounded theory~\cite{nelson2020computational} to identify paradigms through a three-step process. First, we performed cluster analysis on our coded corpus by trying two types of computational methods: KModes clustering and hierarchical clustering. 
For KModes, we combined the elbow method and the silhouette score to compare the clustering results, finding that it performed best at 4 clusters. For hierarchical clustering, we experimented with a set of measures, such as Euclidean distance, Jaccard similarity, and Hamming distance. We used the silhouette score to compare the results and struck a balance between interpretability and clustering performance. The optimal clustering outcomes were achieved with 4 and 6 clusters (Hamming distance). 
Combined with the KModes method, the computational phase yielded three technically-sound clustering results that await further interpretation.

Given the exploratory nature of this study and the inherent subjectivity of the topic, manual intervention was necessary to interpret clusters meaningfully. Therefore, in the second step, we conducted a careful reading of the clusters by identifying the most prominent and exclusive features in each cluster and manually assessed the interpretability of different clustering results. All the authors conducted an anonymous vote, and the results showed unanimous agreement that KModes yielded the most interpretable results. The four clusters it suggested aligned well with our coding experience in the previous section, indicating that affective geovisualization designs may exhibit paradigmatic differences internally.
Finally, we checked the reliability of the clusters using correspondence analysis, projecting the four clusters into a lower-dimensional space to observe their distribution. The visualization revealed relatively clear and distinguishable contours of the four clusters, as seen in \cref{fig:ca}. Detailed analyses and the correspondence analysis biplot are available in the supplementary files.
\begin{figure}[t]
    \centering
    \includegraphics[width=\linewidth]{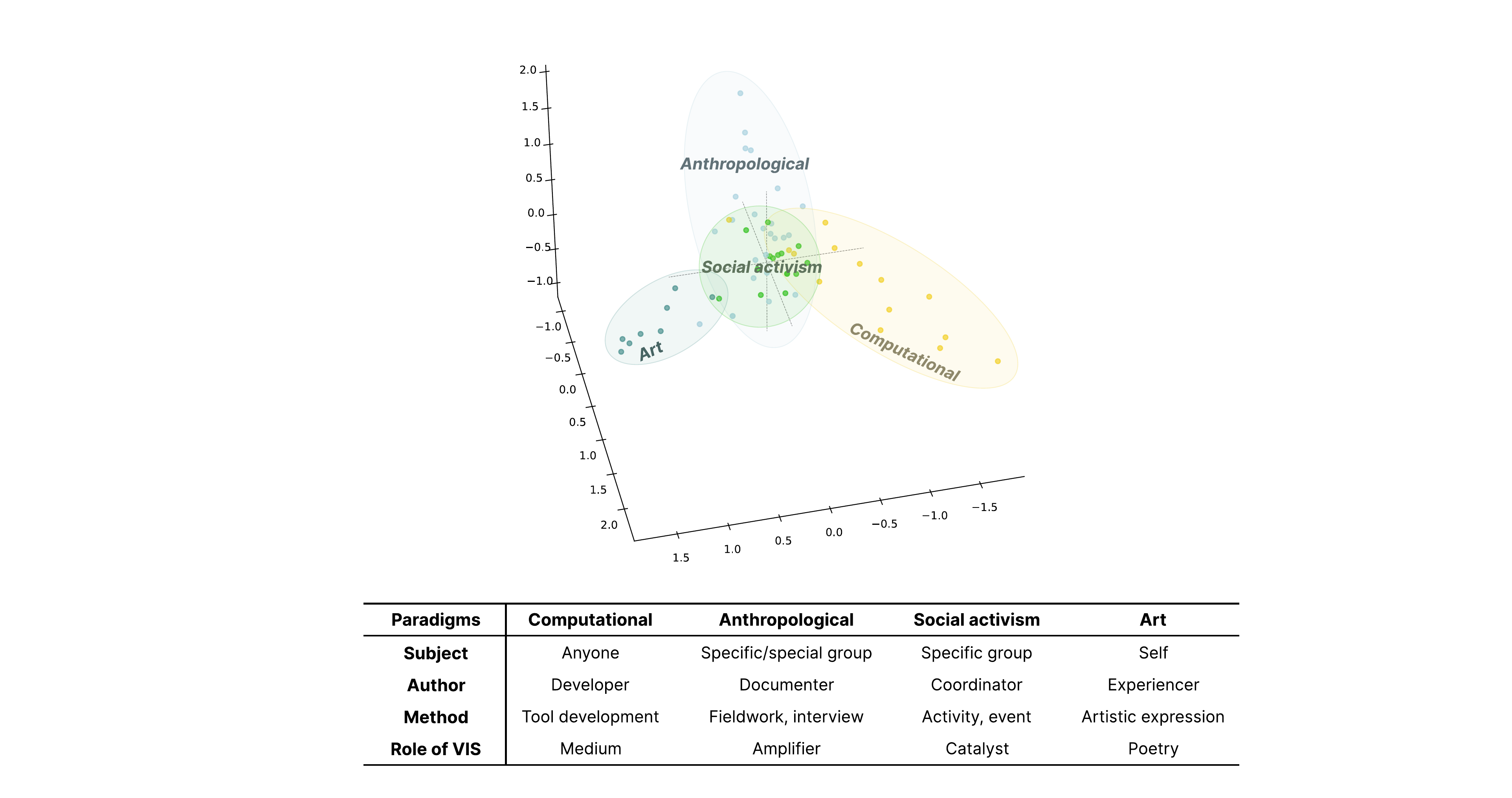}
    \caption{Correspondence Analysis Biplot of the four identified paradigms, along with their characteristics.}
    \label{fig:ca}
    \vspace{-2em}
\end{figure}



\subsection{Patterns}

Based on the interpretation of clustering results, we identified four different design paradigms, distinguished by their methodology, target users, author roles, and visualization functions.

The \textbf{computational paradigm} prioritizes utility, with authors acting as \textit{developers} who create \textit{tools} for the \textit{public}. Here, visualization serves as a \textit{medium} for subjective communication, enabling users to express and share emotions. Emotions can be embedded as direct outputs for social interaction or as quantifiable dimensions for assessing place characteristics. As an example, \textit{Snap Map} (\cref{fig:examples} C) integrates emotional features into its social functions, reflecting a trend toward emotionally aware and user-centered tools. Typical operationalization: Developing tools that collect, process, and visualize emotional data through computational methods such as sentiment analysis.

The \textbf{anthropological paradigm} focuses on giving voice to \textit{specific groups}, particularly vulnerable or underrepresented communities, with authors serving as \textit{documenters}. Visualization functions as an \textit{amplifier}, capturing emotions and translating them into visual narratives. This paradigm often follows an anthropological approach, employing \textit{fieldwork and interviews} to understand emotions. Collaborations with cultural institutions to access historical records are common, such as in \cref{fig:examples} H, where Holocaust testimonies were visualized. The goal is to amplify voices and use visualization for social advocacy.
Typical operationalization: Gathering emotional narratives from specific communities and then translating these into visualizations that help preserve and communicate their own experiences and perspectives.

The \textbf{social activism paradigm} emphasizes the impact of visualization on human emotions. Here, a \textit{specific group}, often open to participation, is invited, and authors take on the role of \textit{coordinators}. This paradigm typically involves \textit{activities/events}, where participants engage with a place firsthand, fostering deeper emotional connections through direct experience. In this context, visualization acts as a \textit{catalyst}, fostering person-place relationships, with emotions emerging as a result of this interaction. For example, as a ``walking data visualization'' exploring sea level rise, \cref{fig:examples} R sparked dialogues and raised awareness of environmental issues among the public.
Typical operationalization: 
Organizing participatory events where visualization supports embodied engagement and collective emotional reflection.

The \textbf{art paradigm} is characterized by personal and artistic expression, with authors acting as \textit{experiencers} who convey their \textit{own} emotion toward places. 
For example, \textit{Grafted Land} (\cref{fig:examples} P) surgically removes disputed areas from the map, covering them with gauze and stitching new borders with green thread. Red-stained gauze reinforces the imagery of wounds and healing, reflecting the areas’ agony and the author's deep emotional connection~\cite{littman2023mapping}.
Here, visualization functions like a \textit{poetry}, expressing sympathy, care or hope for the land.
Typical operationalization: Creating artistic interpretations of place-based emotions through creative visualization techniques.

As shown in \cref{fig:ca}, the computational, art, and anthropological paradigms each stretch along a separate direction, while the social activism paradigm lies at the center, intersecting with the other three to varying degrees. These four clusters interestingly correspond to the general academic categories of science, art, humanities, and social sciences; and social sciences are often viewed as the intermediate bridge between natural science and humanities. Although it is only an exploratory visualization of our limited corpus, this figure intuitively represents the interdisciplinary nature and diverse entry paths of affective geovisualization design.

\section{Discussion}

In this section, we discuss insights from this work and our limitations.

\subsection{Implications for Geographic Applications}

Our analysis of a corpus of affective geovisualization design, based on the PPP model, reveals many possibilities for more human-centered geographic applications.

\textbf{Designing for fine-grained contexts.}
Our analysis of the \textit{person} (\cref{ssec:person}) and \textit{place} (\cref{ssec:place}) dimensions shows that in affective geovisualization design, there is a particularly strong focus on concrete rather than abstract individuals. This includes a wide range of people, from urban citizens and rural indigenous people to transnational migrants and refugees, as well as sexual minorities, the impoverished, and others. These individuals have diverse identities, group sizes, and living spaces, and geographic applications serving them can vary in functions and features. For example, geographic systems designed for groups (\eg families, cultural communities) generally require more sharing and collaboration features on maps. Systems designed for rural and remote areas may need to consider users' knowledge levels and cognitive styles. Like some projects in our corpus, these systems may opt to use more figurative expressions (\eg symbols, drawings) to express geographical experiences, rather than relying too much on abstract thinking (\eg text, data tables). For another example, considering the interpretive levels of places, systems designed for visualizing memorized places need to provide affordances that aid in reminiscing and reconstructing memories. Their visual representations should also account for the missing locations and uncertainty of data.

\textbf{Facilitating subjective geographic experience through diverse methods.}
Regarding the \textit{process} (\cref{ssec:process}) dimension, we have identified rich roles that data and visualization can play in eliciting or augmenting emotion. Among these, some techniques have been discussed more in the visualization community, such as mining emotional information from data and enabling the customization of personal views~\cite{yang2024emogeocity,thudt2015visual}. 
However, we have also discovered many interesting ideas from interdisciplinary literature and practices, such as engaging people through annotating stories on maps and wandering in certain districts. Users are active constructors of human-place relationships and utilize their rich senses (\eg sight, smell, bodily movements) to feel the space. This is quite different from viewing maps as scientific tools, where, for example, a typical task is to optimize the distance between locations and recommend the most efficient route. However, the \textit{dérive} concept frequently seen in affective geovisualization design encourages non-efficiency-driven random exploration. Imagine if a geographic application incorporated this feature; it could help those who wish to adventure, unwind, or break norms achieve a novel and unique geographical experience. Furthermore, as people have rich emotional responses to places, their route planning can be subjective. For instance, they may choose a longer route to avoid locations that have caused them discomfort, or select calming paths and nostalgic routes based on their mood~\cite{klettner2011emomap,bleisch2018exploratory}. 
Considering these user needs and scenarios may lead to the development of more creative and user-friendly applications.

\vspace{-5pt}
\subsection{Reflections on Affective Visualization Design}
As an incremental work in affective visualization design, we have gained a series of reflections from our research:

\textbf{The value of qualitative and mixed methods.}
As \cref{tab:process1} shows, we identified a wide range of methods adopted by affective geovisualization designs to evoke and stimulate emotion, which have not been previously discussed in affective visualization design research. Qualitative methods, in particular, dominate this field. For example, participatory approaches invite people to share personal emotions and experiences, and the embodied approach makes individuals an active part of visualization, allowing them to experience the environment physically.
At the same time, mixed-methods approaches that combine quantitative and qualitative data have also emerged. We find this inspiring because it may address the limitations of purely qualitative methods, such as their small sample sizes, while avoiding the impersonal nature of traditional quantitative methods. For example, \textit{City of Memory} invited people to contribute qualitative data, such as stories about places in the city, on an online platform. Through digital connections, the data, along with its geographical information, can serve both as a valuable archive and as a future dataset~\cite{krygier2006jake}.

\textbf{Inner threads within affective visualization design.}
Affective visualization design is an emerging and complex field. By identifying four distinct design paradigms in \cref{sec:patterns}, we seek to further understand the inner threads within this field. For example, the computational paradigm is suitable for the collection, analysis, and prediction of large-scale data. For instance, by analyzing social media data, the emotions of residents in different urban areas can be visualized to provide data support for urban planning and informing policymakers.
The anthropological paradigm often targets marginalized or vulnerable groups, helping to enhance community members' sense of identity and belonging. For example, in a community project, visualizing the emotional memories of community members can promote community cohesion and cultural heritage. This paradigm can also be applied to historical projects or transformed into cultural tourism resources.
The social activism paradigm uses affective geovisualization tools to display social issues, mobilize public participation, and drive social change, making it likely to be adopted by social activists. For example, visualizing the impact of environmental pollution on residents' emotions can promote and implement environmental protection policies.
The artistic paradigm transforms affective geovisualization data into artistic works. For example, in an exhibition, affective geovisualization works can explore the relationship between humans and nature, evoking emotional resonance among the audience.



\subsection{Limitations and Future Work}
First, due to the lack of available datasets for affective geovisualization, we chose to follow previous research~\cite{lan2023affective} and constructed our own dataset through a combination of literature surveys and snowballing. This method is beneficial for ensuring the feasibility of data searching in a highly interdisciplinary and exploratory study. However, it also means that we were more likely to collect design works related to academia and may have omitted some works that exist in the wild.
Second, our taxonomy is structured based on the PPP model, which proved to be effective in revealing multifaceted interplay between people, place and data visualization. However, we acknowledge that other models or dimensions might offer alternative perspectives.

In terms of future work, we would like to suggest several directions. Technically, researchers can explore the integration of more advanced technologies in affective geovisualization design. This includes the development of automated emotional inference methods, real-time sentiment-aware interfaces, and, as discussed earlier, more fine-grained geographic applications that cater to different user needs and scenario requirements. Theoretically, future research could explore complementary frameworks beyond the PPP model to further the understanding of affective geovisualization design. We also suggest conducting more cross-cultural and longitudinal studies to explore the nuances of how emotional responses to places form, evolve, and vary across different sociocultural contexts. Practically, while our taxonomy and its associated website enable users to browse, search, and filter the identified dimensions and techniques, future work could develop enhanced educational toolkits which translate the framework into more actionable guidelines, such as automated recommendations based on project goals or suggested design templates for different paradigms.


\section{Conclusion}

In this work, we analyzed a corpus of 62 affective geovisualization designs and constructed a taxonomy that characterizes their design features by referring to the PPP model from humanistic geography. Based on the coded corpus, we also identified four high-level paradigms: computational, anthropological, social activism, and art, representing distinct practice patterns of affective geovisualization design in different domains. 
Finally, drawing upon all the findings, we discuss implications for future research and practitioners.

\acknowledgments{
This work was supported by NSFC 62402121 and Shanghai Chenguang Program. We thank all the reviewers for their valuable feedback.}

\bibliographystyle{abbrv-doi-hyperref}
\bibliography{reference}

\begin{thebibliography}{10}

\bibitem{aitken2009into}
S.~Aitken and J.~Craine.
\newblock {Into the image and beyond: Affective visual geographies and GIScience}.
\newblock {\em Qualitative GIS: A mixed methods approach}, pp. 139--55, 2009.

\bibitem{aitken2011affective}
S.~Aitken and J.~Craine.
\newblock {Affective geovisualisations}.
\newblock {\em The Map Reader: Theories of Mapping Practice and Cartographic Representation}, pp. 278--280, 2011. \href{https://doi.org/10.1002/9780470979587.ch36}
{doi: {{%
10\hspace{.1pt}\discretionary{.}{%
}{.}\hspace{.4pt}1002\discretionary{/}{%
}{/}9780470979587\hspace{.1pt}\discretionary{.}{%
}{.}\hspace{.4pt}ch36}}}


\bibitem{al2025unearthing}
A.~Al~Mamun, S.~K. Dey, C.~Zhang, P.~Tiwasing, and O.~Omoloso.
\newblock {Unearthing the multidimensional roles of place attachment in sustainable entrepreneurship: a longitudinal study of an ethnic minority entrepreneur in the UK}.
\newblock {\em International Journal of Entrepreneurial Behavior \& Research}, 2025. \href{https://doi.org/10.1108/IJEBR-11-2023-1146}
{doi: {{%
10\hspace{.1pt}\discretionary{.}{%
}{.}\hspace{.4pt}1108\discretionary{/}{%
}{/}IJEBR\discretionary{%
}{-}{-}11\discretionary{%
}{-}{-}2023\discretionary{%
}{-}{-}1146}}}


\bibitem{anderson2021affective}
C.~L. Anderson and A.~C. Robinson.
\newblock {Affective congruence in visualization design: Influences on reading categorical maps}.
\newblock {\em IEEE Transactions on Visualization and Computer Graphics}, 28(8):2867--2878, 2021. \href{https://doi.org/10.1109/TVCG.2021.3050118}
{doi: {{%
10\hspace{.1pt}\discretionary{.}{%
}{.}\hspace{.4pt}1109\discretionary{/}{%
}{/}TVCG\hspace{.1pt}\discretionary{.}{%
}{.}\hspace{.4pt}2021\hspace{.1pt}\discretionary{.}{%
}{.}\hspace{.4pt}3050118}}}


\bibitem{anderson2001emotional}
K.~Anderson and S.~J. Smith.
\newblock {Emotional geographies}.
\newblock {\em Transactions of the Institute of British geographers}, 26(1):7--10, 2001. \href{https://www.jstor.org/stable/623141}
{doi: {{%
stable\discretionary{/}{%
}{/}623141}}}


\bibitem{andrienko2021visual}
G.~Andrienko, N.~Andrienko, F.~Patterson, S.~Chen, R.~Weibel, H.~Huang, C.~Doulkeridis, H.~Georgiou, N.~Pelekis, Y.~Theodoridis, et~al.
\newblock {Visual analytics for characterizing mobility aspects of urban context}.
\newblock {\em Urban Informatics}, pp. 727--755, 2021. \href{https://doi.org/10.1007/978-981-15-8983-6_40}
{doi: {{%
10\hspace{.1pt}\discretionary{.}{%
}{.}\hspace{.4pt}1007\discretionary{/}{%
}{/}978\discretionary{%
}{-}{-}981\discretionary{%
}{-}{-}15\discretionary{%
}{-}{-}8983\discretionary{%
}{-}{-}6\_40}}}


\bibitem{arksey2005scoping}
H.~Arksey and L.~O'malley.
\newblock {Scoping studies: towards a methodological framework}.
\newblock {\em International journal of social research methodology}, 8(1):19--32, 2005. \href{https://doi.org/10.1080/1364557032000119616}
{doi: {{%
10\hspace{.1pt}\discretionary{.}{%
}{.}\hspace{.4pt}1080\discretionary{/}{%
}{/}1364557032000119616}}}


\bibitem{bartram2017affective}
L.~Bartram, A.~Patra, and M.~Stone.
\newblock {Affective color in visualization}.
\newblock In {\em Proceedings of the CHI Conference on Human Factors in Computing Systems}, pp. 1364--1374, 2017. \href{https://doi.org/10.1145/3025453.3026041}
{doi: {{%
10\hspace{.1pt}\discretionary{.}{%
}{.}\hspace{.4pt}1145\discretionary{/}{%
}{/}3025453\hspace{.1pt}\discretionary{.}{%
}{.}\hspace{.4pt}3026041}}}


\bibitem{bleisch2018exploratory}
S.~Bleisch and D.~Hollenstein.
\newblock {Exploratory geovisualizations for supporting the qualitative analysis and synthesis of place-related emotion data}.
\newblock {\em Cartographic Perspectives}, pp. 30--46, 2018. \href{https://doi.org/10.14714/CP91.1437}
{doi: {{%
10\hspace{.1pt}\discretionary{.}{%
}{.}\hspace{.4pt}14714\discretionary{/}{%
}{/}CP91\hspace{.1pt}\discretionary{.}{%
}{.}\hspace{.4pt}1437}}}


\bibitem{bondi2017making}
L.~Bondi.
\newblock {Making connections and thinking through emotions: between geography and psychotherapy}.
\newblock {\em Transactions of the Institute of British Geographers}, 30(4):433--448, 2005. \href{https://doi.org/10.1111/j.1475-5661.2005.00183.x}
{doi: {{%
10\hspace{.1pt}\discretionary{.}{%
}{.}\hspace{.4pt}1111\discretionary{/}{%
}{/}j\hspace{.1pt}\discretionary{.}{%
}{.}\hspace{.4pt}1475\discretionary{%
}{-}{-}5661\hspace{.1pt}\discretionary{.}{%
}{.}\hspace{.4pt}2005\hspace{.1pt}\discretionary{.}{%
}{.}\hspace{.4pt}00183\hspace{.1pt}\discretionary{.}{%
}{.}\hspace{.4pt}x}}}


\bibitem{bondi2016emotional}
L.~Bondi.
\newblock {\em {Emotional geographies}}.
\newblock Routledge, 2016.

\bibitem{boy2017showing}
J.~Boy, A.~V. Pandey, J.~Emerson, M.~Satterthwaite, O.~Nov, and E.~Bertini.
\newblock {Showing people behind data: Does anthropomorphizing visualizations elicit more empathy for human rights data?}
\newblock In {\em Proceedings of the CHI Conference on Human Factors in Computing Systems}, pp. 5462--5474, 2017. \href{https://doi.org/10.1145/3025453.3025512}
{doi: {{%
10\hspace{.1pt}\discretionary{.}{%
}{.}\hspace{.4pt}1145\discretionary{/}{%
}{/}3025453\hspace{.1pt}\discretionary{.}{%
}{.}\hspace{.4pt}3025512}}}


\bibitem{bruno2002atlas}
G.~Bruno.
\newblock {\em {Atlas of emotion: Journeys in art, architecture, and film}}.
\newblock Verso, 2002.

\bibitem{burns2023we}
A.~Burns, C.~Lee, R.~Chawla, E.~Peck, and N.~Mahyar.
\newblock {Who do we mean when we talk about visualization novices?}
\newblock In {\em Proceedings of the CHI Conference on Human Factors in Computing Systems}, pp. 1--16, 2023. \href{https://doi.org/10.1145/3544548.3581524}
{doi: {{%
10\hspace{.1pt}\discretionary{.}{%
}{.}\hspace{.4pt}1145\discretionary{/}{%
}{/}3544548\hspace{.1pt}\discretionary{.}{%
}{.}\hspace{.4pt}3581524}}}


\bibitem{cao2012whisper}
N.~Cao, Y.-R. Lin, X.~Sun, D.~Lazer, S.~Liu, and H.~Qu.
\newblock {Whisper: Tracing the spatiotemporal process of information diffusion in real time}.
\newblock {\em IEEE Transactions on Visualization and Computer Graphics}, 18(12):2649--2658, 2012. \href{https://doi.org/10.1109/TVCG.2012.291}
{doi: {{%
10\hspace{.1pt}\discretionary{.}{%
}{.}\hspace{.4pt}1109\discretionary{/}{%
}{/}TVCG\hspace{.1pt}\discretionary{.}{%
}{.}\hspace{.4pt}2012\hspace{.1pt}\discretionary{.}{%
}{.}\hspace{.4pt}291}}}


\bibitem{carpendale2017subjectivity}
S.~Carpendale, A.~Thudt, C.~Perin, and W.~Willett.
\newblock {Subjectivity in personal storytelling with visualization}.
\newblock {\em Information Design Journal}, 23(1):48--64, 2017. \href{https://doi.org/10.1145/3544548.3581524}
{doi: {{%
10\hspace{.1pt}\discretionary{.}{%
}{.}\hspace{.4pt}1145\discretionary{/}{%
}{/}3544548\hspace{.1pt}\discretionary{.}{%
}{.}\hspace{.4pt}3581524}}}


\bibitem{cartwright2008developing}
W.~Cartwright, A.~Miles, B.~Morris, L.~Vaughan, and J.~Yuille.
\newblock {Developing concepts for an affective atlas}.
\newblock {\em Springer Berlin Heidelberg}, pp. 219--234, 2008.

\bibitem{chen2023sensemap}
J.~Chen, Q.~Huang, C.~Wang, and C.~Li.
\newblock {SenseMap: Urban performance visualization and analytics via semantic textual similarity}.
\newblock {\em IEEE Transactions on Visualization and Computer Graphics}, 2023. \href{https://doi.org/10.1109/TVCG.2023.3333356}
{doi: {{%
10\hspace{.1pt}\discretionary{.}{%
}{.}\hspace{.4pt}1109\discretionary{/}{%
}{/}TVCG\hspace{.1pt}\discretionary{.}{%
}{.}\hspace{.4pt}2023\hspace{.1pt}\discretionary{.}{%
}{.}\hspace{.4pt}3333356}}}


\bibitem{chen2017vaud}
W.~Chen, Z.~Huang, F.~Wu, M.~Zhu, H.~Guan, and R.~Maciejewski.
\newblock {VAUD: A visual analysis approach for exploring spatio-temporal urban data}.
\newblock {\em IEEE Transactions on Visualization and Computer Graphics}, 24(9):2636--2648, 2017. \href{https://doi.org/10.1109/TVCG.2017.2758362}
{doi: {{%
10\hspace{.1pt}\discretionary{.}{%
}{.}\hspace{.4pt}1109\discretionary{/}{%
}{/}TVCG\hspace{.1pt}\discretionary{.}{%
}{.}\hspace{.4pt}2017\hspace{.1pt}\discretionary{.}{%
}{.}\hspace{.4pt}2758362}}}


\bibitem{cho2015vairoma}
I.~Cho, W.~Dou, D.~X. Wang, E.~Sauda, and W.~Ribarsky.
\newblock {Vairoma: A visual analytics system for making sense of places, times, and events in roman history}.
\newblock {\em IEEE Transactions on Visualization and Computer Graphics}, 22(1):210--219, 2015. \href{https://doi.org/10.1109/TVCG.2015.2467971}
{doi: {{%
10\hspace{.1pt}\discretionary{.}{%
}{.}\hspace{.4pt}1109\discretionary{/}{%
}{/}TVCG\hspace{.1pt}\discretionary{.}{%
}{.}\hspace{.4pt}2015\hspace{.1pt}\discretionary{.}{%
}{.}\hspace{.4pt}2467971}}}


\bibitem{counted2016making}
V.~Counted.
\newblock {Making sense of place attachment: Towards a holistic understanding of people-place relationships and experiences}.
\newblock {\em Environment, space, place}, 8(1):7--32, 2016. \href{https://doi.org/10.5840/esplace2016811}
{doi: {{%
10\hspace{.1pt}\discretionary{.}{%
}{.}\hspace{.4pt}5840\discretionary{/}{%
}{/}esplace2016811}}}


\bibitem{davidson2017phobic}
J.~Davidson.
\newblock {\em {Phobic geographies: The phenomenology and spatiality of identity}}.
\newblock Routledge, 2017. \href{https://doi.org/10.4324/9781315246864}
{doi: {{%
10\hspace{.1pt}\discretionary{.}{%
}{.}\hspace{.4pt}4324\discretionary{/}{%
}{/}9781315246864}}}


\bibitem{davidson2004embodying}
J.~Davidson and C.~Milligan.
\newblock {Embodying emotion sensing space: introducing emotional geographies}, 2004. \href{https://doi.org/10.1080/1464936042000317677}
{doi: {{%
10\hspace{.1pt}\discretionary{.}{%
}{.}\hspace{.4pt}1080\discretionary{/}{%
}{/}1464936042000317677}}}


\bibitem{debord1958theory}
G.~Debord.
\newblock {Theory of the D{\'e}rive}.
\newblock {\em Visual Culture: Spaces of visual culture}, 3:77--82, 1958.

\bibitem{elo2008qualitative}
S.~Elo and H.~Kyng{\"a}s.
\newblock {The qualitative content analysis process}.
\newblock {\em Journal of advanced nursing}, 62(1):107--115, 2008. \href{https://doi.org/10.1111/j.1365-2648.2007.04569.x}
{doi: {{%
10\hspace{.1pt}\discretionary{.}{%
}{.}\hspace{.4pt}1111\discretionary{/}{%
}{/}j\hspace{.1pt}\discretionary{.}{%
}{.}\hspace{.4pt}1365\discretionary{%
}{-}{-}2648\hspace{.1pt}\discretionary{.}{%
}{.}\hspace{.4pt}2007\hspace{.1pt}\discretionary{.}{%
}{.}\hspace{.4pt}04569\hspace{.1pt}\discretionary{.}{%
}{.}\hspace{.4pt}x}}}


\bibitem{fathullah2018engaging}
A.~Fathullah and K.~S. Willis.
\newblock {Engaging the senses: The potential of emotional data for participation in urban planning}.
\newblock {\em Urban Science}, 2(4):98, 2018. \href{https://doi.org/10.3390/urbansci2040098}
{doi: {{%
10\hspace{.1pt}\discretionary{.}{%
}{.}\hspace{.4pt}3390\discretionary{/}{%
}{/}urbansci2040098}}}


\bibitem{feng2022survey}
Z.~Feng, H.~Qu, S.-H. Yang, Y.~Ding, and J.~Song.
\newblock {A survey of visual analytics in urban area}.
\newblock {\em Expert Systems}, 39(9):e13065, 2022. \href{https://doi.org/10.1111/exsy.13065}
{doi: {{%
10\hspace{.1pt}\discretionary{.}{%
}{.}\hspace{.4pt}1111\discretionary{/}{%
}{/}exsy\hspace{.1pt}\discretionary{.}{%
}{.}\hspace{.4pt}13065}}}


\bibitem{friendly2001milestones}
M.~Friendly and D.~J. Denis.
\newblock {Milestones in the history of thematic cartography, statistical graphics, and data visualization}.
\newblock \url{http://www.datavis.ca/milestones}, 2001.

\bibitem{fu2023more}
Y.~Fu and J.~Stasko.
\newblock {More than data stories: Broadening the role of visualization in contemporary journalism}.
\newblock {\em IEEE Transactions on Visualization and Computer Graphics}, 2023. \href{https://doi.org/10.1109/TVCG.2023.3287585}
{doi: {{%
10\hspace{.1pt}\discretionary{.}{%
}{.}\hspace{.4pt}1109\discretionary{/}{%
}{/}TVCG\hspace{.1pt}\discretionary{.}{%
}{.}\hspace{.4pt}2023\hspace{.1pt}\discretionary{.}{%
}{.}\hspace{.4pt}3287585}}}


\bibitem{gardener2017walk}
J.~Gardener.
\newblock {A Walk in the Park: Perceptions of Place through Mapping}.
\newblock {\em Cartographic Perspectives}, 88:34--38, 2017.

\bibitem{gardener2018interdisciplinary}
J.~Gardener, W.~Cartwright, and L.~Duxbury.
\newblock {An interdisciplinary approach to mapping through scientific cartography, design and artistic expression}.
\newblock In {\em Proceedings of the ICA}, vol.~1, pp. 1--6. Copernicus GmbH, 2018. \href{https://doi.org/10.5194/ica-proc-1-43-2018}
{doi: {{%
10\hspace{.1pt}\discretionary{.}{%
}{.}\hspace{.4pt}5194\discretionary{/}{%
}{/}ica\discretionary{%
}{-}{-}proc\discretionary{%
}{-}{-}1\discretionary{%
}{-}{-}43\discretionary{%
}{-}{-}2018}}}


\bibitem{garreton2023attitudinal}
M.~Garret{\'o}n, F.~Morini, P.~Celhay, M.~D{\"o}rk, and D.~Parra.
\newblock {Attitudinal effects of data visualizations and illustrations in data stories}.
\newblock {\em IEEE Transactions on Visualization and Computer Graphics}, 2023. \href{https://doi.org/10.1109/TVCG.2023.3248319}
{doi: {{%
10\hspace{.1pt}\discretionary{.}{%
}{.}\hspace{.4pt}1109\discretionary{/}{%
}{/}TVCG\hspace{.1pt}\discretionary{.}{%
}{.}\hspace{.4pt}2023\hspace{.1pt}\discretionary{.}{%
}{.}\hspace{.4pt}3248319}}}


\bibitem{gilmartin1991effects}
P.~Gilmartin and R.~Lloyd.
\newblock {The effects of map projections and map distance on emotional involvement with places}.
\newblock {\em The Cartographic Journal}, 28(2):145--151, 1991. \href{https://www.tandfonline.com/doi/abs/10.1179/000870491787859241}
{doi: {{%
doi\discretionary{/}{%
}{/}abs\discretionary{/}{%
}{/}10\hspace{.1pt}\discretionary{.}{%
}{.}\hspace{.4pt}1179\discretionary{/}{%
}{/}000870491787859241}}}


\bibitem{golbiowska2020rainbow}
I.~M. Go{\l}biowska and A.~{\c{C}}{\"o}ltekin.
\newblock {Rainbow dash: Intuitiveness, interpretability and memorability of the rainbow color scheme in visualization}.
\newblock {\em IEEE Transactions on Visualization and Computer Graphics}, 28(7):2722--2733, 2020. \href{https://doi.org/10.1109/TVCG.2020.3035823}
{doi: {{%
10\hspace{.1pt}\discretionary{.}{%
}{.}\hspace{.4pt}1109\discretionary{/}{%
}{/}TVCG\hspace{.1pt}\discretionary{.}{%
}{.}\hspace{.4pt}2020\hspace{.1pt}\discretionary{.}{%
}{.}\hspace{.4pt}3035823}}}


\bibitem{guo2023liberroad}
Y.~Guo, Y.~Luo, K.~Lu, L.~Li, H.~Yang, and X.~Yuan.
\newblock {Liberroad: Probing into the journey of chinese classics through visual analytics}.
\newblock {\em IEEE Transactions on Visualization and Computer Graphics}, 2023. \href{https://doi.org/10.1109/TVCG.2023.3326944}
{doi: {{%
10\hspace{.1pt}\discretionary{.}{%
}{.}\hspace{.4pt}1109\discretionary{/}{%
}{/}TVCG\hspace{.1pt}\discretionary{.}{%
}{.}\hspace{.4pt}2023\hspace{.1pt}\discretionary{.}{%
}{.}\hspace{.4pt}3326944}}}


\bibitem{harley1989deconstructing}
J.~B. Harley.
\newblock {Deconstructing the map}.
\newblock {\em Cartographica: The international journal for geographic information and geovisualization}, 26(2):1--20, 1989. \href{https://doi.org/10.3138/E635-7827-1757-9T53}
{doi: {{%
10\hspace{.1pt}\discretionary{.}{%
}{.}\hspace{.4pt}3138\discretionary{/}{%
}{/}E635\discretionary{%
}{-}{-}7827\discretionary{%
}{-}{-}1757\discretionary{%
}{-}{-}9T53}}}


\bibitem{harley2008maps}
J.~B. Harley.
\newblock {Maps, knowledge, and power}.
\newblock In {\em Geographic Thought}, pp. 129--148. Routledge, 2008. \href{https://doi.org/10.4324/9780203893074}
{doi: {{%
10\hspace{.1pt}\discretionary{.}{%
}{.}\hspace{.4pt}4324\discretionary{/}{%
}{/}9780203893074}}}


\bibitem{harrower2003colorbrewer}
M.~Harrower and C.~A. Brewer.
\newblock {ColorBrewer.org: an online tool for selecting colour schemes for maps}.
\newblock {\em The Cartographic Journal}, 40(1):27--37, 2003. \href{https://www.tandfonline.com/doi/abs/10.1179/000870403235002042}
{doi: {{%
doi\discretionary{/}{%
}{/}abs\discretionary{/}{%
}{/}10\hspace{.1pt}\discretionary{.}{%
}{.}\hspace{.4pt}1179\discretionary{/}{%
}{/}000870403235002042}}}


\bibitem{hay1998sense}
R.~Hay.
\newblock {Sense of place in developmental context}.
\newblock {\em Journal of environmental psychology}, 18(1):5--29, 1998. \href{https://doi.org/10.1006/jevp.1997.0060}
{doi: {{%
10\hspace{.1pt}\discretionary{.}{%
}{.}\hspace{.4pt}1006\discretionary{/}{%
}{/}jevp\hspace{.1pt}\discretionary{.}{%
}{.}\hspace{.4pt}1997\hspace{.1pt}\discretionary{.}{%
}{.}\hspace{.4pt}0060}}}


\bibitem{he2019interactive}
T.~He, J.~Bao, S.~Ruan, R.~Li, Y.~Li, H.~He, and Y.~Zheng.
\newblock {Interactive bike lane planning using sharing bikes’ trajectories}.
\newblock {\em IEEE Transactions on Knowledge and Data Engineering}, 32(8):1529--1542, 2019. \href{https://doi.org/10.1109/TKDE.2019.2907091}
{doi: {{%
10\hspace{.1pt}\discretionary{.}{%
}{.}\hspace{.4pt}1109\discretionary{/}{%
}{/}TKDE\hspace{.1pt}\discretionary{.}{%
}{.}\hspace{.4pt}2019\hspace{.1pt}\discretionary{.}{%
}{.}\hspace{.4pt}2907091}}}


\bibitem{iturrioz2010artistic}
T.~Iturrioz and M.~Wachowicz.
\newblock {An artistic perspective for affective cartography}.
\newblock In {\em Mapping different geographies}, pp. 74--92. Springer, 2010. \href{https://doi.org/10.1007/978-3-642-15537-6_5}
{doi: {{%
10\hspace{.1pt}\discretionary{.}{%
}{.}\hspace{.4pt}1007\discretionary{/}{%
}{/}978\discretionary{%
}{-}{-}3\discretionary{%
}{-}{-}642\discretionary{%
}{-}{-}15537\discretionary{%
}{-}{-}6\_5}}}


\bibitem{kelly2016collectively}
M.~Kelly.
\newblock {Collectively mapping borders}.
\newblock {\em Cartographic Perspectives}, (84):31--38, 2016. \href{https://doi.org/10.14714/CP84.1363}
{doi: {{%
10\hspace{.1pt}\discretionary{.}{%
}{.}\hspace{.4pt}14714\discretionary{/}{%
}{/}CP84\hspace{.1pt}\discretionary{.}{%
}{.}\hspace{.4pt}1363}}}


\bibitem{kennedy2018feeling}
H.~Kennedy and R.~L. Hill.
\newblock {The feeling of numbers: Emotions in everyday engagements with data and their visualisation}.
\newblock {\em Sociology}, 52(4):830--848, 2018. \href{https://doi.org/10.1177/0038038516674675}
{doi: {{%
10\hspace{.1pt}\discretionary{.}{%
}{.}\hspace{.4pt}1177\discretionary{/}{%
}{/}0038038516674675}}}


\bibitem{klettner2011emomap}
S.~Klettner, H.~Huang, and M.~Schmidt.
\newblock {EmoMap - considering emotional responses to space for enhancing LBS}.
\newblock In {\em International Symposium on Location-Based Services}, pp. 300--303, 2011. \href{https://www.researchgate.net/publication/320801034}
{doi: {{%
publication\discretionary{/}{%
}{/}320801034}}}


\bibitem{knowles2015inductive}
A.~K. Knowles, L.~Westerveld, and L.~Strom.
\newblock {Inductive visualization: A humanistic alternative to GIS}.
\newblock {\em GeoHumanities}, 1(2):233--265, 2015. \href{https://doi.org/10.1080/2373566X.2015.1108831}
{doi: {{%
10\hspace{.1pt}\discretionary{.}{%
}{.}\hspace{.4pt}1080\discretionary{/}{%
}{/}2373566X\hspace{.1pt}\discretionary{.}{%
}{.}\hspace{.4pt}2015\hspace{.1pt}\discretionary{.}{%
}{.}\hspace{.4pt}1108831}}}


\bibitem{krygier2006jake}
J.~Krygier.
\newblock {Jake Barton’s performance maps: An essay}.
\newblock {\em Cartographic perspectives}, (53):41--50, 2006. \href{https://doi.org/10.14714/CP53.361}
{doi: {{%
10\hspace{.1pt}\discretionary{.}{%
}{.}\hspace{.4pt}14714\discretionary{/}{%
}{/}CP53\hspace{.1pt}\discretionary{.}{%
}{.}\hspace{.4pt}361}}}


\bibitem{lan2021kineticharts}
X.~Lan, Y.~Shi, Y.~Wu, X.~Jiao, and N.~Cao.
\newblock {Kineticharts: Augmenting affective expressiveness of charts in data stories with animation design}.
\newblock {\em IEEE Transactions on Visualization and Computer Graphics}, 28(1):933--943, 2021. \href{https://doi.org/10.1109/TVCG.2021.3114775}
{doi: {{%
10\hspace{.1pt}\discretionary{.}{%
}{.}\hspace{.4pt}1109\discretionary{/}{%
}{/}TVCG\hspace{.1pt}\discretionary{.}{%
}{.}\hspace{.4pt}2021\hspace{.1pt}\discretionary{.}{%
}{.}\hspace{.4pt}3114775}}}


\bibitem{lan2023affective}
X.~Lan, Y.~Wu, and N.~Cao.
\newblock {Affective visualization design: Leveraging the emotional impact of data}.
\newblock {\em IEEE Transactions on Visualization and Computer Graphics}, 2024. \href{https://doi.org/10.1109/TVCG.2023.3327385}
{doi: {{%
10\hspace{.1pt}\discretionary{.}{%
}{.}\hspace{.4pt}1109\discretionary{/}{%
}{/}TVCG\hspace{.1pt}\discretionary{.}{%
}{.}\hspace{.4pt}2023\hspace{.1pt}\discretionary{.}{%
}{.}\hspace{.4pt}3327385}}}


\bibitem{lan2022negative}
X.~Lan, Y.~Wu, Y.~Shi, Q.~Chen, and N.~Cao.
\newblock {Negative emotions, positive outcomes? exploring the communication of negativity in serious data stories}.
\newblock In {\em Proceedings of the CHI Conference on Human Factors in Computing Systems}, pp. 1--14, 2022. \href{https://doi.org/10.1145/3491102.3517530}
{doi: {{%
10\hspace{.1pt}\discretionary{.}{%
}{.}\hspace{.4pt}1145\discretionary{/}{%
}{/}3491102\hspace{.1pt}\discretionary{.}{%
}{.}\hspace{.4pt}3517530}}}


\bibitem{latif2021deeper}
S.~Latif, S.~Chen, and F.~Beck.
\newblock {A Deeper Understanding of Visualization-Text Interplay in Geographic Data-driven Stories}.
\newblock In {\em Computer Graphics Forum}, pp. 311--322, 2021. \href{https://doi.org/10.1111/cgf.14309}
{doi: {{%
10\hspace{.1pt}\discretionary{.}{%
}{.}\hspace{.4pt}1111\discretionary{/}{%
}{/}cgf\hspace{.1pt}\discretionary{.}{%
}{.}\hspace{.4pt}14309}}}


\bibitem{lee2022affective}
E.~Lee-Robbins and E.~Adar.
\newblock {Affective learning objectives for communicative visualizations}.
\newblock {\em IEEE Transactions on Visualization and Computer Graphics}, 29(1):1--11, 2022. \href{https://doi.org/10.1109/TVCG.2022.3209500}
{doi: {{%
10\hspace{.1pt}\discretionary{.}{%
}{.}\hspace{.4pt}1109\discretionary{/}{%
}{/}TVCG\hspace{.1pt}\discretionary{.}{%
}{.}\hspace{.4pt}2022\hspace{.1pt}\discretionary{.}{%
}{.}\hspace{.4pt}3209500}}}


\bibitem{ley2014humanistic}
D.~Ley and M.~Samuels.
\newblock {\em {Humanistic Geography (RLE Social \& Cultural Geography): Problems and Prospects}}.
\newblock Routledge, 2014. \href{https://doi.org/10.4324/9781315819655}
{doi: {{%
10\hspace{.1pt}\discretionary{.}{%
}{.}\hspace{.4pt}4324\discretionary{/}{%
}{/}9781315819655}}}


\bibitem{li2020warehouse}
Q.~Li, Q.~Liu, C.~Tang, Z.~Li, S.~Wei, X.~Peng, M.~Zheng, T.~Chen, and Q.~Yang.
\newblock {Warehouse vis: A visual analytics approach to facilitating warehouse location selection for business districts}.
\newblock In {\em Computer graphics forum}, pp. 483--495, 2020. \href{https://doi.org/10.1111/cgf.13996}
{doi: {{%
10\hspace{.1pt}\discretionary{.}{%
}{.}\hspace{.4pt}1111\discretionary{/}{%
}{/}cgf\hspace{.1pt}\discretionary{.}{%
}{.}\hspace{.4pt}13996}}}


\bibitem{li2023geocamera}
W.~Li, Z.~Wang, Y.~Wang, D.~Weng, L.~Xie, S.~Chen, H.~Zhang, and H.~Qu.
\newblock {GeoCamera: Telling stories in geographic visualizations with camera movements}.
\newblock In {\em Proceedings of the CHI Conference on Human Factors in Computing Systems}, pp. 1--15, 2023. \href{https://doi.org/10.1145/3544548.358147}
{doi: {{%
10\hspace{.1pt}\discretionary{.}{%
}{.}\hspace{.4pt}1145\discretionary{/}{%
}{/}3544548\hspace{.1pt}\discretionary{.}{%
}{.}\hspace{.4pt}358147}}}


\bibitem{littman2023mapping}
A.~Littman.
\newblock {Mapping the Wound: Feminine Gestures of Empathy and Healing}.
\newblock {\em Borders in Globalization Review}, 4(2):72--87, 2023. \href{https://doi.org/10.18357/bigr42202321513}
{doi: {{%
10\hspace{.1pt}\discretionary{.}{%
}{.}\hspace{.4pt}18357\discretionary{/}{%
}{/}bigr42202321513}}}


\bibitem{long2020ai}
D.~Long and B.~Magerko.
\newblock {What is AI literacy? Competencies and design considerations}.
\newblock In {\em Proceedings of the CHI Conference on Human Factors in Computing Systems}, pp. 1--16, 2020. \href{https://doi.org/10.1145/3313831.3376727}
{doi: {{%
10\hspace{.1pt}\discretionary{.}{%
}{.}\hspace{.4pt}1145\discretionary{/}{%
}{/}3313831\hspace{.1pt}\discretionary{.}{%
}{.}\hspace{.4pt}3376727}}}


\bibitem{mayer2023characterization}
B.~Mayer, N.~Steinhauer, B.~Preim, and M.~Meuschke.
\newblock {A Characterization of Interactive Visual Data Stories With a Spatio-Temporal Context}.
\newblock In {\em Computer Graphics Forum}, vol.~42, p. e14922. Wiley Online Library, 2023. \href{https://doi.org/10.1111/cgf.14922}
{doi: {{%
10\hspace{.1pt}\discretionary{.}{%
}{.}\hspace{.4pt}1111\discretionary{/}{%
}{/}cgf\hspace{.1pt}\discretionary{.}{%
}{.}\hspace{.4pt}14922}}}


\bibitem{mclean2020temporalities}
K.~McLean.
\newblock {Temporalities of the smellscape: Creative mapping as visual representation}.
\newblock {\em Modern approaches to the visualization of landscapes}, pp. 217--245, 2020. \href{https://doi.org/10.1007/978-3-658-30956-5_12}
{doi: {{%
10\hspace{.1pt}\discretionary{.}{%
}{.}\hspace{.4pt}1007\discretionary{/}{%
}{/}978\discretionary{%
}{-}{-}3\discretionary{%
}{-}{-}658\discretionary{%
}{-}{-}30956\discretionary{%
}{-}{-}5\_12}}}


\bibitem{mclean2019nose}
K.~J. McLean.
\newblock {\em {Nose-first: practices of smellwalking and smellscape mapping}}.
\newblock Royal College of Art (United Kingdom), 2019.

\bibitem{milligan2016healing}
C.~Milligan, A.~Bingley, and A.~Gatrell.
\newblock {‘Healing and Feeling': The Place of Emotions in Later Life}.
\newblock In {\em Emotional geographies}, pp. 49--62. Routledge, 2016. \href{https://doi.org/10.4324/9781315579245}
{doi: {{%
10\hspace{.1pt}\discretionary{.}{%
}{.}\hspace{.4pt}4324\discretionary{/}{%
}{/}9781315579245}}}


\bibitem{moro2012peripatetic}
S.~Moro.
\newblock {Peripatetic box and personal mapping: From studio to classroom to city}.
\newblock {\em Mapping Cultures: Place, Practice, Performance}, pp. 260--279, 2012. \href{https://doi.org/10.1057/9781137025050_14}
{doi: {{%
10\hspace{.1pt}\discretionary{.}{%
}{.}\hspace{.4pt}1057\discretionary{/}{%
}{/}9781137025050\_14}}}


\bibitem{moro2021mapping}
S.~Moro.
\newblock {\em {Mapping paradigms in modern and contemporary art: poetic cartography}}.
\newblock Routledge, 2021.

\bibitem{muehlenhaus2012if}
I.~Muehlenhaus.
\newblock {If looks could kill: The impact of different rhetorical styles on persuasive geocommunication}.
\newblock {\em The Cartographic Journal}, 49(4):361--375, 2012. \href{https://doi.org/10.1179/1743277412Y.0000000032}
{doi: {{%
10\hspace{.1pt}\discretionary{.}{%
}{.}\hspace{.4pt}1179\discretionary{/}{%
}{/}1743277412Y\hspace{.1pt}\discretionary{.}{%
}{.}\hspace{.4pt}0000000032}}}


\bibitem{nast1998places}
H.~J. Nast, S.~Pile, and H.~J. Nast.
\newblock {\em {Places through the body}}.
\newblock Routledge London, 1998.

\bibitem{nelson2020computational}
L.~K. Nelson.
\newblock {Computational grounded theory: A methodological framework}.
\newblock {\em Sociological Methods \& Research}, 49(1):3--42, 2020. \href{https://doi.org/10.1177/0049124117729703}
{doi: {{%
10\hspace{.1pt}\discretionary{.}{%
}{.}\hspace{.4pt}1177\discretionary{/}{%
}{/}0049124117729703}}}


\bibitem{nold2009emotional}
C.~Nold.
\newblock {\em {Emotional cartography: Technologies of the self}}.
\newblock Space Studios, 2009.

\bibitem{nold2018bio}
C.~Nold.
\newblock {Bio Mapping: How can we use emotion to articulate cities?}
\newblock {\em Livingmaps Review}, 4, 2018.

\bibitem{nusrat2016evaluating}
S.~Nusrat, M.~J. Alam, and S.~Kobourov.
\newblock {Evaluating cartogram effectiveness}.
\newblock {\em IEEE Transactions on Visualization and Computer Graphics}, 24(2):1077--1090, 2016. \href{https://doi.org/10.1109/TVCG.2016.2642109}
{doi: {{%
10\hspace{.1pt}\discretionary{.}{%
}{.}\hspace{.4pt}1109\discretionary{/}{%
}{/}TVCG\hspace{.1pt}\discretionary{.}{%
}{.}\hspace{.4pt}2016\hspace{.1pt}\discretionary{.}{%
}{.}\hspace{.4pt}2642109}}}


\bibitem{parr1999delusional}
H.~Parr.
\newblock {Delusional geographies: the experiential worlds of people during madness/illness}.
\newblock {\em Environment and Planning D: Society and Space}, 17(6):673--690, 1999. \href{https://doi.org/10.1068/d170673}
{doi: {{%
10\hspace{.1pt}\discretionary{.}{%
}{.}\hspace{.4pt}1068\discretionary{/}{%
}{/}d170673}}}


\bibitem{pearce2010mapping}
M.~W. Pearce and M.~J. Hermann.
\newblock {Mapping Champlain's travels: Restorative techniques for historical cartography}.
\newblock {\em Cartographica: The International Journal for Geographic Information and Geovisualization}, 45(1):32--46, 2010. \href{https://doi.org/10.3138/carto.45.1.32}
{doi: {{%
10\hspace{.1pt}\discretionary{.}{%
}{.}\hspace{.4pt}3138\discretionary{/}{%
}{/}carto\hspace{.1pt}\discretionary{.}{%
}{.}\hspace{.4pt}45\hspace{.1pt}\discretionary{.}{%
}{.}\hspace{.4pt}1\hspace{.1pt}\discretionary{.}{%
}{.}\hspace{.4pt}32}}}


\bibitem{pinder1996subverting}
D.~Pinder.
\newblock {Subverting cartography: The situationists and maps of the city}.
\newblock {\em Environment and planning A}, 28(3):405--427, 1996. \href{https://doi.org/10.1068/a280405}
{doi: {{%
10\hspace{.1pt}\discretionary{.}{%
}{.}\hspace{.4pt}1068\discretionary{/}{%
}{/}a280405}}}


\bibitem{powell2010making}
K.~Powell.
\newblock {Making sense of place: Mapping as a multisensory research method}.
\newblock {\em Qualitative Inquiry}, 16(7):539--555, 2010. \href{https://doi.org/10.1177/1077800410372600}
{doi: {{%
10\hspace{.1pt}\discretionary{.}{%
}{.}\hspace{.4pt}1177\discretionary{/}{%
}{/}1077800410372600}}}


\bibitem{qu2007visual}
H.~Qu, W.-Y. Chan, A.~Xu, K.-L. Chung, K.-H. Lau, and P.~Guo.
\newblock {Visual analysis of the air pollution problem in Hong Kong}.
\newblock {\em IEEE Transactions on visualization and Computer Graphics}, 13(6):1408--1415, 2007. \href{https://doi.org/10.1109/TVCG.2007.70523}
{doi: {{%
10\hspace{.1pt}\discretionary{.}{%
}{.}\hspace{.4pt}1109\discretionary{/}{%
}{/}TVCG\hspace{.1pt}\discretionary{.}{%
}{.}\hspace{.4pt}2007\hspace{.1pt}\discretionary{.}{%
}{.}\hspace{.4pt}70523}}}


\bibitem{roth2013empirically}
R.~E. Roth.
\newblock {An empirically-derived taxonomy of interaction primitives for interactive cartography and geovisualization}.
\newblock {\em IEEE Transactions on Visualization and Computer Graphics}, 19(12):2356--2365, 2013. \href{https://doi.org/10.1109/TVCG.2013.130}
{doi: {{%
10\hspace{.1pt}\discretionary{.}{%
}{.}\hspace{.4pt}1109\discretionary{/}{%
}{/}TVCG\hspace{.1pt}\discretionary{.}{%
}{.}\hspace{.4pt}2013\hspace{.1pt}\discretionary{.}{%
}{.}\hspace{.4pt}130}}}


\bibitem{roth2021cartographic}
R.~E. Roth.
\newblock {Cartographic design as visual storytelling: synthesis and review of map-based narratives, genres, and tropes}.
\newblock {\em The Cartographic Journal}, 58(1):83--114, 2021. \href{https://doi.org/10.1080/00087041.2019.1633103}
{doi: {{%
10\hspace{.1pt}\discretionary{.}{%
}{.}\hspace{.4pt}1080\discretionary{/}{%
}{/}00087041\hspace{.1pt}\discretionary{.}{%
}{.}\hspace{.4pt}2019\hspace{.1pt}\discretionary{.}{%
}{.}\hspace{.4pt}1633103}}}


\bibitem{scannell2010defining}
L.~Scannell and R.~Gifford.
\newblock {Defining place attachment: A tripartite organizing framework}.
\newblock {\em Journal of environmental psychology}, 30(1):1--10, 2010. \href{https://doi.org/10.1016/j.jenvp.2009.09.006}
{doi: {{%
10\hspace{.1pt}\discretionary{.}{%
}{.}\hspace{.4pt}1016\discretionary{/}{%
}{/}j\hspace{.1pt}\discretionary{.}{%
}{.}\hspace{.4pt}jenvp\hspace{.1pt}\discretionary{.}{%
}{.}\hspace{.4pt}2009\hspace{.1pt}\discretionary{.}{%
}{.}\hspace{.4pt}09\hspace{.1pt}\discretionary{.}{%
}{.}\hspace{.4pt}006}}}


\bibitem{shen2017streetvizor}
Q.~Shen, W.~Zeng, Y.~Ye, S.~M. Arisona, S.~Schubiger, R.~Burkhard, and H.~Qu.
\newblock {StreetVizor: Visual exploration of human-scale urban forms based on street views}.
\newblock {\em IEEE Transactions on Visualization and Computer Graphics}, 24(1):1004--1013, 2017. \href{https://doi.org/10.1109/TVCG.2017.2744159}
{doi: {{%
10\hspace{.1pt}\discretionary{.}{%
}{.}\hspace{.4pt}1109\discretionary{/}{%
}{/}TVCG\hspace{.1pt}\discretionary{.}{%
}{.}\hspace{.4pt}2017\hspace{.1pt}\discretionary{.}{%
}{.}\hspace{.4pt}2744159}}}


\bibitem{shi2022breaking}
Y.~Shi, T.~Gao, X.~Jiao, and N.~Cao.
\newblock {Breaking the fourth wall of data stories through interaction}.
\newblock {\em IEEE Transactions on Visualization and Computer Graphics}, 29(1):972--982, 2022. \href{https://doi.org/10.1109/TVCG.2022.3209409}
{doi: {{%
10\hspace{.1pt}\discretionary{.}{%
}{.}\hspace{.4pt}1109\discretionary{/}{%
}{/}TVCG\hspace{.1pt}\discretionary{.}{%
}{.}\hspace{.4pt}2022\hspace{.1pt}\discretionary{.}{%
}{.}\hspace{.4pt}3209409}}}


\bibitem{slavik2024using}
C.~E. Slavik, C.~Fish, and E.~Peters.
\newblock {Using Geovisualizations to Educate the Public About Environmental Health Hazards: What Works and Why}.
\newblock {\em Current Environmental Health Reports}, 11(4):453--467, 2024. \href{https://doi.org/10.1007/s40572-024-00461-8}
{doi: {{%
10\hspace{.1pt}\discretionary{.}{%
}{.}\hspace{.4pt}1007\discretionary{/}{%
}{/}s40572\discretionary{%
}{-}{-}024\discretionary{%
}{-}{-}00461\discretionary{%
}{-}{-}8}}}


\bibitem{smith2009emotion}
M.~Smith, L.~Bondi, M.~Smith, and J.~Davidson.
\newblock {\em {Emotion, place and culture}}, vol.~1.
\newblock Ashgate Farnham, 2009.

\bibitem{thudt2015visual}
A.~Thudt, D.~Baur, S.~Huron, and S.~Carpendale.
\newblock {Visual mementos: Reflecting memories with personal data}.
\newblock {\em IEEE Transactions on Visualization and Computer Graphics}, 22(1):369--378, 2015. \href{https://doi.org/10.1109/TVCG.2015.2467831}
{doi: {{%
10\hspace{.1pt}\discretionary{.}{%
}{.}\hspace{.4pt}1109\discretionary{/}{%
}{/}TVCG\hspace{.1pt}\discretionary{.}{%
}{.}\hspace{.4pt}2015\hspace{.1pt}\discretionary{.}{%
}{.}\hspace{.4pt}2467831}}}


\bibitem{thudt2018self}
A.~Thudt, U.~Hinrichs, S.~Huron, and S.~Carpendale.
\newblock {Self-reflection and personal physicalization construction}.
\newblock In {\em Proceedings of the CHI Conference on Human Factors in Computing Systems}, pp. 1--13, 2018. \href{https://doi.org/10.1145/3173574.3173728}
{doi: {{%
10\hspace{.1pt}\discretionary{.}{%
}{.}\hspace{.4pt}1145\discretionary{/}{%
}{/}3173574\hspace{.1pt}\discretionary{.}{%
}{.}\hspace{.4pt}3173728}}}


\bibitem{tuan1975place}
Y.-F. Tuan.
\newblock {Place: an experiential perspective}.
\newblock {\em Geographical review}, pp. 151--165, 1975. \href{https://www.jstor.org/stable/213970}
{doi: {{%
stable\discretionary{/}{%
}{/}213970}}}


\bibitem{tuan2017humanistic}
Y.-F. Tuan.
\newblock {Humanistic geography}.
\newblock In {\em Theory and Methods}, pp. 127--138. Routledge, 2017. \href{https://doi.org/10.4324/9781315236285}
{doi: {{%
10\hspace{.1pt}\discretionary{.}{%
}{.}\hspace{.4pt}4324\discretionary{/}{%
}{/}9781315236285}}}


\bibitem{wang2019emotional}
Y.~Wang, A.~Segal, R.~Klatzky, D.~F. Keefe, P.~Isenberg, J.~Hurtienne, E.~Hornecker, T.~Dwyer, and S.~Barrass.
\newblock {An emotional response to the value of visualization}.
\newblock {\em IEEE computer graphics and applications}, 39(5):8--17, 2019. \href{https://doi.org/10.1109/MCG.2019.2923483}
{doi: {{%
10\hspace{.1pt}\discretionary{.}{%
}{.}\hspace{.4pt}1109\discretionary{/}{%
}{/}MCG\hspace{.1pt}\discretionary{.}{%
}{.}\hspace{.4pt}2019\hspace{.1pt}\discretionary{.}{%
}{.}\hspace{.4pt}2923483}}}


\bibitem{weng2020towards}
D.~Weng, C.~Zheng, Z.~Deng, M.~Ma, J.~Bao, Y.~Zheng, M.~Xu, and Y.~Wu.
\newblock {Towards better bus networks: A visual analytics approach}.
\newblock {\em IEEE Transactions on Visualization and Computer Graphics}, 27(2):817--827, 2020. \href{https://doi.org/10.1109/TVCG.2020.3030458}
{doi: {{%
10\hspace{.1pt}\discretionary{.}{%
}{.}\hspace{.4pt}1109\discretionary{/}{%
}{/}TVCG\hspace{.1pt}\discretionary{.}{%
}{.}\hspace{.4pt}2020\hspace{.1pt}\discretionary{.}{%
}{.}\hspace{.4pt}3030458}}}


\bibitem{wood1992power}
D.~Wood and J.~Fels.
\newblock {\em {The power of maps}}.
\newblock Guilford Press, 1992.

\bibitem{wu2017exploring}
H.~Wu, H.~Fan, and S.~Wu.
\newblock {Exploring spatiotemporal patterns of long-distance taxi rides in Shanghai}.
\newblock {\em ISPRS International Journal of Geo-Information}, 6(11):339, 2017. \href{https://doi.org/10.3390/ijgi6110339}
{doi: {{%
10\hspace{.1pt}\discretionary{.}{%
}{.}\hspace{.4pt}3390\discretionary{/}{%
}{/}ijgi6110339}}}


\bibitem{yang2024emogeocity}
Y.~Yang, Y.~Liu, Q.~Bai, T.~Zhou, Z.~Ye, and X.~Dong.
\newblock {EmoGeoCity: Interactive Visual Exploration of City’s Historical and Cultural Evolution Based on Emotional Geography}.
\newblock In {\em IEEE Pacific Visualization Conference}, pp. 102--111. IEEE, 2024. \href{https://doi.org/10.1109/PacificVis60374.2024.00020}
{doi: {{%
10\hspace{.1pt}\discretionary{.}{%
}{.}\hspace{.4pt}1109\discretionary{/}{%
}{/}PacificVis60374\hspace{.1pt}\discretionary{.}{%
}{.}\hspace{.4pt}2024\hspace{.1pt}\discretionary{.}{%
}{.}\hspace{.4pt}00020}}}


\bibitem{young2013spatial}
J.~C. Young and M.~P. Gilmore.
\newblock {The spatial politics of affect and emotion in participatory GIS}.
\newblock {\em Annals of the Association of American Geographers}, 103(4):808--823, 2013. \href{https://doi.org/10.1080/00045608.2012.707596}
{doi: {{%
10\hspace{.1pt}\discretionary{.}{%
}{.}\hspace{.4pt}1080\discretionary{/}{%
}{/}00045608\hspace{.1pt}\discretionary{.}{%
}{.}\hspace{.4pt}2012\hspace{.1pt}\discretionary{.}{%
}{.}\hspace{.4pt}707596}}}


\end{thebibliography}

\end{document}